\documentclass[twocolumn]{aastex631}

\usepackage{amsmath}
\usepackage{orcidlink}
\usepackage{bm}
\usepackage{xcolor}  
\DeclareMathOperator*{\E}{\mathbb{E}}

\makeatletter
\patchcmd{\acknowledgments}
  {\begin{internallinenumbers}}
  {\ifnumlines\begin{internallinenumbers}\fi}{}{}
\patchcmd{\endacknowledgments}
  {\end{internallinenumbers}}
  {\ifnumlines\end{internallinenumbers}\fi}{}{}
\patchcmd{\acknowledgements}
  {\begin{internallinenumbers}}
  {\ifnumlines\begin{internallinenumbers}\fi}{}{}
\patchcmd{\endacknowledgements}
  {\end{internallinenumbers}}
  {\ifnumlines\end{internallinenumbers}\fi}{}{}
\makeatother

\begin{document}

\title{POSTELLAR: Posterior Stellar Spectrum Sampling - An Alternative to Approximate Stellar Spectra for Exoplanetary Analysis}

\correspondingauthor{Dhvani Doshi}
\email{dhvani.doshi@mail.mcgill.ca}

\author[0000-0003-3610-3434]{Dhvani Doshi}
\affiliation{Department of Physics, McGill University, 3600 Rue University, Montréal, QC H3A 2T8, Canada}
\affiliation{Trottier Space Institute, McGill University, 3550 Rue University, Montr\'eal, QC H3A 2A7, Canada}
\affiliation{Ciela Institute, Montréal, Canada}

\author[0000-0001-6129-5699]{Nicolas B. Cowan}
\affiliation{Department of Physics, McGill University, 3600 Rue University, Montréal, QC H3A 2T8, Canada}
\affiliation{Trottier Space Institute, McGill University, 3550 Rue University, Montr\'eal, QC H3A 2A7, Canada}
\affiliation{Department of Earth and Planetary Sciences, McGill University, 3450 Rue University, Montréal, QC H3A 0E8, Canada}

\author[0000-0002-8669-5733]{Yashar Hezaveh}
\affiliation{Département de Physique, Université de Montréal, Montréal QC, Canada}
\affiliation{Mila—Quebec Artificial Intelligence Institute, Montréal, QC, Canada}
\affiliation{Ciela Institute, Montréal, Canada}

\author[0009-0008-5839-5937]{Gabriel Missael Barco}
\affiliation{Département de Physique, Université de Montréal, Montréal QC, Canada}
\affiliation{Mila—Quebec Artificial Intelligence Institute, Montréal, QC, Canada}
\affiliation{Ciela Institute, Montréal, Canada}

\author[0000-0003-3506-5667]{Étienne Artigau}
\affiliation{Département de Physique, Université de Montréal, Montréal QC, Canada}
\affiliation{Institut Trottier de Recherche sur les Exoplanètes, Université de Montréal, 1375 Avenue Thérèse-Lavoie-Roux, Montréal, QC H2V 0B3, Canada}
\affiliation{Observatoire du Mont-Mégantic, Université de Montréal,
Montréal H3C 3J7, Canada}

\begin{abstract}
We present a novel approach to perform posterior sampling of the underlying stellar spectrum in high-resolution spectroscopic observations. Our method, \texttt{postellar}, coherently combines information from empirical observations and physics-based models, enabling more accurate spectral recovery while providing uncertainties that can be propagated into downstream analyses. This is accomplished by treating the intrinsic stellar spectrum as a latent variable and performing posterior sampling under a Gaussian likelihood with an informative prior constructed using a score-based diffusion model trained on PHOENIX stellar models. We validate the framework on synthetic SPIRou radial velocity (RV) observations generated from the empirical spectra of Barnard’s Star and Proxima Centauri. Spectra inferred with \texttt{postellar} more accurately recover the ground truth than standard empirical templates. For medium signal-to-noise observations, \texttt{postellar} improves RV accuracy by up to a factor of three, while RVs derived from empirical templates tend to be biased. This method performs particularly well in low signal-to-noise and low-cadence regimes. The postellar framework is broadly applicable to other high-resolution spectroscopy science cases including stellar abundance analyses and exoplanet atmospheric characterization.
\end{abstract}

\section{Introduction} \label{sec:intro}
Exoplanet characterization is fundamentally a differential measurement where planetary properties are inferred from changes imprinted on the host star's observed light. As a result, our ability to detect and characterize planets is first-order limited by how well we understand the stellar spectrum and its intrinsic variability. Inadequate knowledge of the stellar baseline or time-dependent activity directly propagates into biased planetary inferences~\citep{cameron_impact_2024}.

This limitation is relevant for most exoplanet characterization techniques. For example, in transmission spectroscopy and photometry, retrieving planetary signals requires an accurate stellar baseline to separate the stellar and planetary contributions to the observed light. Similar challenges arise for emission spectroscopy, which requires the stellar spectral energy distribution to be known reliably to determine the planet–star flux contrast and infer the planet's atmospheric temperature–pressure structures~\citep{madhusudhan_temperature_2009}. Stellar spectra are also central to constraining planetary composition and formation through host-star abundances, which in turn depend on accurate stellar modelling and high-quality spectral characterization~\citep{lebzelter_comparative_2012,teske_starplanet_2024}.

In practice, defining an accurate stellar baseline is in part challenging because stars are intrinsically time-variable. Magnetic activity produces spots, flares, and granulation that alter spectral features on timescales ranging from minutes to years~\citep{burt_precise_2025}. Consequently, a substantial body of work has focused on modelling or mitigating stellar activity to improve exoplanet detection and characterization. These efforts include Gaussian process models of correlated stellar variability~\citep[e.g.,][]{GP}, activity indicators derived from chromospheric-sensitive spectral lines~\citep[e.g.,][]{activity}, and physically motivated models of active stellar surfaces \citep[e.g.,][]{Cristo}. While these approaches aim to separate planetary signals from stellar variability, they generally assume that the underlying stellar spectrum has already been measured accurately. Our work addresses this complementary problem: recovering the underlying stellar spectrum itself.

Knowledge of the stellar spectrum is crucially important for radial velocity (RV) analysis. Here, planetary masses are inferred from stellar radial velocities by measuring Doppler shifts in the stellar spectrum induced by the planet’s gravitational pull. The planetary induced shifts are extremely small, much smaller than a pixel even at high spectral resolution, and must be disentangled from much larger apparent shifts caused by the Earth’s barycentric motion. Accurately correcting for this motion and reliably registering spectra across epochs therefore requires very precise knowledge of the intrinsic, unshifted stellar spectrum~\citep{dumusque_planetary_2011,burt_precise_2025}. 

Early RV methods relied on line lists for cross-correlation~\citep{griffin_photoelectric_1967, baranne_coravel_1979, pepe_coralie_2002}. While effective for FGK stars, these approaches do not probe the full RV content for M dwarfs, whose spectra are dominated by dense, blended molecular features that cannot be accurately reproduced by current stellar models~\citep{lafarga_carmenes_2020,rainer_stellar_2020}. As a result, empirical template-matching methods were introduced, in which high-SNR stellar spectra are constructed by co-adding observations. This approach underpins many widely used RV pipelines, including HARPS-TERRA~\citep{anglada-escude_harps-terra_2012}, NAIRA~\citep{astudillo-defru_harps_2017-1}, SERVAL~\citep{Zechmeister2018}, S-BART~\citep{silva_novel_2022}, and APERO~\citep{cook_apero_2022}, as well as iterative template-reconstruction techniques~\citep{gao_retrieval_2016} and line-by-line RV extraction methods~\citep{artigau_line-by-line_2022}.

While empirical templates capture the full RV information content of the stellar spectrum, they also imprint observational noise into the model by construction. When measuring RVs using derivatives of the template, the derivative determination is even more prone to the noise than first-order flux measurements. Moreover, stellar atmosphere models or line lists, though physically motivated, remain insufficiently accurate across the full spectral range~\citep{Jahandar2024}. Motivated by these limitations, we introduce a forward-modelling approach that retains the flexibility and completeness of empirical templates while avoiding the direct incorporation of observational noise~\citep{hogg_frizzle_2024}.

We propose a forward-modelling approach, similar to~\cite{bedell}'s \texttt{wobble}, in which the intrinsic stellar spectrum is treated as a latent variable and explicitly sampled from the data. By enabling posterior sampling of the stellar spectrum, we generate spectral realizations that are informed by physically motivated priors derived from stellar models, while the likelihood enforces sufficient flexibility to accurately reproduce the observed spectra. This framework effectively unifies the strengths of traditional line-list–based models and empirical templates, offering a complete and continuous spectral model.

\section{Methodology} \label{sec:method}
\subsection{Outline}
In this framework, we treat the stellar spectrum as a variable and sample from the joint probability distribution $P(f, v \mid Y)$, which describes the probability of the spectrum, $f$, and radial velocity, $v$, given the data, $Y$. In traditional techniques, the stellar spectrum is fixed to a line list or empirical spectrum, therefore we only sample from the conditional probability, $P(v \mid Y,f)$. In practice however, $f$, is not known. Conditioning on a fixed value of $f$ therefore collapses the joint posterior, which may lead to biased RV estimates and underestimated uncertainties. Therefore, we sample from the spectral conditional posterior distribution $p(f \mid Y, v)$ and use these spectra to sample from $p(v \mid Y, f)$.

To sample $p(f \mid Y, v)$, we express the log posterior using Bayes' theorem, which conveniently separates the likelihood, prior, and evidence terms. To avoid approximating the evidence, we focus on the gradient of the log posterior with respect to the spectrum, $f$, commonly referred to as the score of the log posterior. This gives

\begin{equation}
    \nabla_{f} \log p(f \mid Y, v) = \nabla_{f} \log p(Y \mid f, v) + \nabla_{f} \log p(f), \label{eq:bayes}
\end{equation}

where the first term represents the score of the log likelihood. Here, the likelihood anchors the inference to the data and is modelled as a multivariate Gaussian, with its width set by the observational uncertainties. Since $f$ is a continuous variable, we introduce a forward model $A(f, v)$ that maps the parameters $f$ and $v$ into the observation (pixel) space, producing a model prediction with the same dimensionality as the data vector $Y$. The forward model is discussed further in Section~\ref{sec:forward}. 

Defining an informative prior on $f$ is also essential, as it encodes the probability of spectral features, correlations between features, and the shapes of individual lines. We construct this prior by training a neural network (discussed further in Section~\ref{sec:training}), specifically a score-based model, to learn the distribution of M-dwarf spectra from PHOENIX models~\citep{Song2021sde}. Then by coupling this learned prior and the likelihood function with a stochastic differential equation, we can generate posterior samples of the underlying stellar spectrum consistent with the observations (discussed further in Section~\ref{sec:prior} and Section~\ref{sec:postspecsample}). Given posterior samples of the spectrum, we then sample $p(v \mid Y, f)$ using the method described in Section~\ref{sec:rvsampling} to retrieve RVs from the same data.

\subsection{Forward Model}
\label{sec:forward}
We develop an RV retrieval methodology based on the principle that the underlying stellar spectrum in RV observations is an unknown variable. We first describe how an intrinsic stellar spectrum is forward-modelled to produce the observed data from a given instrument, thereby outlining the components that contribute to an RV observation. This process is described by
\begin{equation}
    {Y} = {A}({f},v) + {\eta} 
    \label{eq:data}
\end{equation}
where ${Y}$ is a 2D vector representing the observed spectrum, $Y(\lambda)$, measured over $N_{\mathrm{obs}}$ observations by a given instrument. The intrinsic stellar spectrum is denoted by $f$, which can also be expressed as $f(\lambda)$. The function $A(f,v)$ represents the forward-modelling process and the noise model is captured by $\eta$. The forward model depends on the radial velocity, $v$, of the star induced by the planet. Here, $v$ is a vector of length $N_{\mathrm{obs}}$, with each element corresponding to the planetary induced radial velocity for a particular observation.

To enable subsequent posterior sampling, we restrict the forward model to a linear regime. Consequently, we do not incorporate the effects of instrumental broadening, rotational broadening nor time-varying stellar processes such as stellar activity in this framework. While Doppler shifting is not inherently linear, it can be well approximated as a linear operation in the regime of small planetary RVs considered here. This allows for a matrix representation of the forward-modelling process where ${A}({f},v) \approx \tilde{A}_v f$ and $\tilde{A}_v$ is the Jacobian of the forward map with respect to $f$, evaluated at $v$. We will henceforth refer to it as $\tilde{A}$. If the RV is large enough to be problematic, one can update the BERV value and/or Doppler shift the wavelength grid and sample for the local RV to linearize the problem.

Under these assumptions, we consider ${f}$ to be defined at the instrument's spectral resolution. This can be justified because, at the sensitivity relevant for our analysis, convolution and Doppler shifting are commutable operations. We further assume that ${f}$ does not exhibit signatures of rotational broadening or stellar activity. More specifically, we assume that locally on a pixel scale, any change due to activity is vanishingly small (after removing RV jitter) compared to noise.

With these assumptions, the forward-modelling process is captured by
\begin{equation}
    A({f},v) = D(\lambda_{\text{obs}}, S(\lambda_{\text{int}}, v_{\text{sys}},v_{\text{BERV}},f, v)),
    \label{eq:forwardmodel}
\end{equation}
where $S$ is the Doppler shift operator acting on the intrinsic wavelength grid $\lambda_{\text{int}}$, where we include the assumed radial velocity of the system, $v_{\text{sys}}$, and the Barycentric Earth Radial Velocity (BERV), $v_{\text{BERV}}$. We define $D$ as the downsampling operator that maps the spectrum onto the observed wavelength grid $\lambda_{\text{obs}}$. Consequently, this formulation assumes that the instrument wavelength solution and line spread function are fixed.

Lastly, ${\eta}$ represents additive noise in the observations. In this work, we consider only photon noise and assume a regime in which it can be approximated as Gaussian. Therefore, our forward model begins with an intrinsic stellar spectrum $f$ at the spectral resolution of the instrument, which undergoes a Doppler shift by total total velocity, $v_{\text{sys}}+v_{\text{BERV}}+v$, and then subsequent downsampling. The observed signal is then obtained by the addition of Gaussian photon noise.

\subsection{Spectrum Sampling Framework}
\label{sec:specsample}
The prior distribution of $f$ is inherently high-dimensional, as $f=f(\lambda)$ assigns a flux value to each wavelength bin. It encodes both the correlations between wavelength bins and the overall probability of how a stellar spectrum should appear across wavelengths. Using a line list collapses this prior to a delta function, which is overly constraining and likely misspecified. An empirical template, in contrast, imposes a uniform prior, assuming no prior knowledge of spectral features and relying entirely on the data. This discards known information about absorption lines and general line shapes, limiting the ability to leverage existing spectral knowledge.

Several approaches have been proposed to approximate the spectral prior using Gaussian distributions. For example,~\cite{bedell} applies L2 regularization, which places a Gaussian prior on parameters that are then optimized to fit the spectrum. Others model the spectrum itself using Gaussian processes, or represent it as a Gaussian process mixture model of spectra from other stars~\citep{czekala_disentangling_2017,rajpaul_robust_2020,feeney_ssspang_2021}. However, as seen in ~\cite{feeney_ssspang_2021}, the prior distributions of stellar spectra are highly non-Gaussian.

To leverage known spectral features while avoiding a heavily misspecified prior, we require a method to characterize the full, high-dimensional prior. Generative machine learning models provide different ways of learning a distribution $p_{\theta}$ from a dataset $\{f^{(i)}\}_{i=1}^N$, allowing one to either sample new data points $f \sim p_{\theta}$ or evaluate the density $p_{\theta}(f)$.

There are different approaches, including normalizing flows~\citep[e.g.,][]{Papamakarios2017maf}, score-based models~\citep[e.g.,][]{Song2021sde}, and variational autoencoders~\citep{Kingma2014vae}, among others. These kinds of generative models can be used to learn complex, data-driven prior distributions from observational data or simulations and provide a principled way to incorporate known characteristics of the parameters of interest. Once a generative prior is learned, it can be used to perform posterior inference~\citep{chung2023dps, Whang2021nf_inverse}. In astrophysics, they have been applied to, for example, strong gravitational lensing source inversion~\citep[e.g.,][]{Adam2022}, and exoplanetary atmospheric retrieval~\citep[e.g.,][]{Vasist2023exoplanet_nf}.

In particular, Score-based models (SBMs) provide one such framework, excelling at modelling high-dimensional distributions~\citep{Dhariwal2021sbm_vs_gans, Peebles2022ScalableDM}. An SBM is a machine learning model in which a neural network, $s_{\theta}(f, t)$, is trained to learn the score of the log probability distribution, $\nabla_{f_t}\log{p_t(f_t)}$, convolved with different levels of noise, indexed by $t$, as further described in the following section. Once this model has been trained sufficiently over a set of spectra, it would be able to sample new spectra from the approximate true prior distribution.

\subsubsection{Training the Prior Model}
\label{sec:training}
In a typical supervised machine learning setup, training the model would require access to the true score of the distribution. However, if this quantity was available, there would be no need for an SBM to be trained in the first place. To overcome this, we employ the denoising score matching (DSM) objective ~\citep{Hyvarinen2005, Vincent2011}, which allows the model to learn the score of a distribution without direct access to the true score $\nabla_{f_t}\log p_t(f_t)$.


In this framework, our model, $s_\theta(f,t)$ takes as input a spectrum $f$ and a ``time'' perturbation parameter $t\sim \mathcal{U}(0, 1)$, where we use the notation $t$ to represent a continuous progression from unperturbed ($t=0$) to fully perturbed ($t=1$) samples, analogous to a temporal annealing process. We anneal the input spectrum such that the perturbed sample is defined as
\begin{equation}
    f_t = \mu(t)f_0+\sigma(t)z,
    \label{eq:anneal}
\end{equation}
with z being drawn from a multivariate normal distribution with mean zero and covariance equal to the identity matrix ($z \sim \mathcal{N}(0, \mathbf{I})$) and $f_0$ is a spectrum from our training dataset, $\mathcal{D}$. The functions $\mu(t)$ and $\sigma(t)$ (defined in Section~\ref{sec:prior}) control the level of perturbation, where $t=0$ means there is no Gaussian noise added, while $t=1$ corresponds to complete Gaussian noise. This procedure effectively anneals the samples and the overall distribution across perturbation levels. 

Consequently, the model outputs the score of the log prior at different perturbation levels $\nabla_{f_t}\log{p_t(f_t)}$ such that
\begin{equation}
    \nabla_{f_t}\log p_t(f_t) = \nabla_{f_t}\log \int p_{\mathcal{D}}(f)p_t(f_t \mid f)df,
\end{equation}
where $p_t(f_t \mid f)$ is a perturbation kernel, typically Gaussian $\mathcal{N}(\mu(t), \sigma(t))$, and $p_{\mathcal{D}}(f)$ is the true data distribution.

Traditionally, the loss function, which is used to update the model parameters during the training process, is the expected squared error between the model output and the true score:
\begin{equation}
    L_{\theta} = \E_{\substack{
f_0 \sim \mathcal{D} \\
t \sim \mathcal{U}(0,1) \\
f_t \sim p(f_t \mid f_0)
}} \Big[\space||s_\theta(f_t,t) - \nabla_{f_t}\log{p_t(f_t)}||^2\space\Big].
\label{eq:lossfn}
\end{equation}

However, \citet{Hyvarinen2005} and \citet{Vincent2011} show that this function can be rewritten so that it does not require access to the true score $\nabla_{f_t} \log p_t(f_t)$. Specifically, minimizing the expected squared error between the model score and the conditional score $\nabla_{f_t} \log p(f_t \mid f_0)$ yields the same optimal parameters (up to an additive constant independent of the model). From Equation~\ref{eq:anneal}, we find that $\nabla_{f_t}\log{p(f_t \mid f_0)} = \frac{\mu(t)f_0 - f_t}{\sigma(t)^2}$, thus we can write the loss function with known quantities while only requiring samples from the true data distribution $f_0 \sim p_\mathcal{D}$.

This allows us to define a tractable loss function and train $s_\theta(f,t)$. We train a neural network with a U-net architecture~\citep{Ronneberger2015unet} using the \texttt{score-model} package from~\cite{Adam2022}. Through training, the model learns an approximation of  $\nabla_{f_t}\log{p_t(f_t)}$, and in particular at $t=0$ we recover an approximation of the score of the true prior distribution, which is the quantity of interest. 

Typically, one would use this score to sample the underlying distribution. However, as discussed earlier, this distribution is multimodal and highly complex. While SBMs are efficient for high-dimensional distributions, they typically cannot approximate complex distributions well, partly because low-probability regions are underrepresented in the training set $\mathcal{D}$, leaving the model with insufficient data to learn these regions accurately. 

However, SBMs are particularly well suited to approximating smooth distributions. At $t=1$, the perturbed distribution is Gaussian, which the model can represent accurately and stably. We exploit this property by initializing samples from the $t=1$ distribution and progressively propagating them toward $t=0$, thereby transforming them into samples approximately from the true prior. This progressive denoising can be viewed as a form of annealed sampling, in which one gradually transitions from a simple, tractable distribution to a complex target distribution. Similar strategies are employed in Monte Carlo methods, when sampling from multimodal distributions, even when the analytic form of the density is known. By evolving samples through this sequence of intermediate distributions, we can reliably approximate the high-dimensional prior, including regions where the training data are sparse, leading to a faithful approximation of the underlying spectral distribution.

\subsubsection{Sampling the Prior Distribution}
\label{sec:prior}
In the previous section, we trained a SBM across multiple perturbation levels. Our objective is to propagate samples $f_t$ across these levels while preserving the underlying probability distribution, so that we can ultimately generate samples from the true distribution. Using a discrete set of noise scales is computationally inefficient, since the system would fall out of equilibrium each time the noise level changes.

To address this, we instead adopt a continuous noise scale and model the perturbation process using stochastic differential equations (SDEs). Specifically, we use the variance-preserving (VP) SDE~\citep{Ho2020ddpm, Song2021sde}, which describes how noise is gradually added to the samples through a combination of drift and diffusion:
\begin{equation}
    df_t = -\frac{1}{2}\beta(t)f_tdt + g(t)dw_t.
    \label{eq:forsde}
\end{equation}
Here, $f_t$ is given by Equation~\ref{eq:anneal} and is perturbed by $df_t$ under an infinitesimal change $dt$ in the perturbation level. The noise schedule is defined as $\beta(t)=\beta_{\text{min}}+t(\beta_{\text{max}}-\beta_{\text{min}})$ where $\beta_{\text{min}}=10^{-2}$ and $\beta_{\text{max}}=20$ and $g(t)=\sqrt{\beta(t)}$. As a result, from Equation~\ref{eq:anneal}, we find that $\mu(t)=\text{exp}(-\frac{1}{2}\int^t_0\beta(s)ds)$ and $\sigma(t)=[1-\text{exp}(-\int^t_0\beta(s)ds)]^{1/2}$. 

The function $\beta(t)$ controls both the strength of the drift and the magnitude of the diffusion. Finally, $dw_t$ is the Wiener process, adding stochastic Brownian motion to the samples. Together, these dynamics describe how samples evolve from low to high perturbation levels.

The evolution of the probability density $p_t(f_t)$ under this SDE is governed by the Fokker--Planck equation, which characterizes how probability mass changes over time. In particular, the diffusion term tends to spread probability mass, while the drift term induces a contraction of the distribution. This framework allows us to define a reverse-time process that transports samples back toward the data distribution. To transport samples from the high-perturbation regime back to the true distribution, we use the reverse-time SDE derived with Anderson’s formula,
\begin{equation}
\begin{aligned}
    df_t &= \left[-\frac{1}{2}\beta(t)f_t-g^2(t)\nabla_{f_t}\log{p_t(f_t)}\right]d\overline{t} \\
    &\quad + g(t)d\overline{w_t},
\end{aligned}
\label{eq:revsde}
\end{equation}
where $d\bar{t}$ denotes an infinitesimal step backward in time and $d\bar{w}_t$ is the corresponding time-reversed Wiener process. The score function $\nabla_{f_t} \log p_t(f_t)$ is approximated by the trained SBM, denoted $s_\theta(f_t, t)$.

This reverse process guides the samples back by their drift movement and towards higher-density regions using the score while retaining stochasticity. In practice, we solve this reverse SDE numerically using the Euler--Maruyama method.

\begin{figure*}
    \centering
    \includegraphics[width=\linewidth]{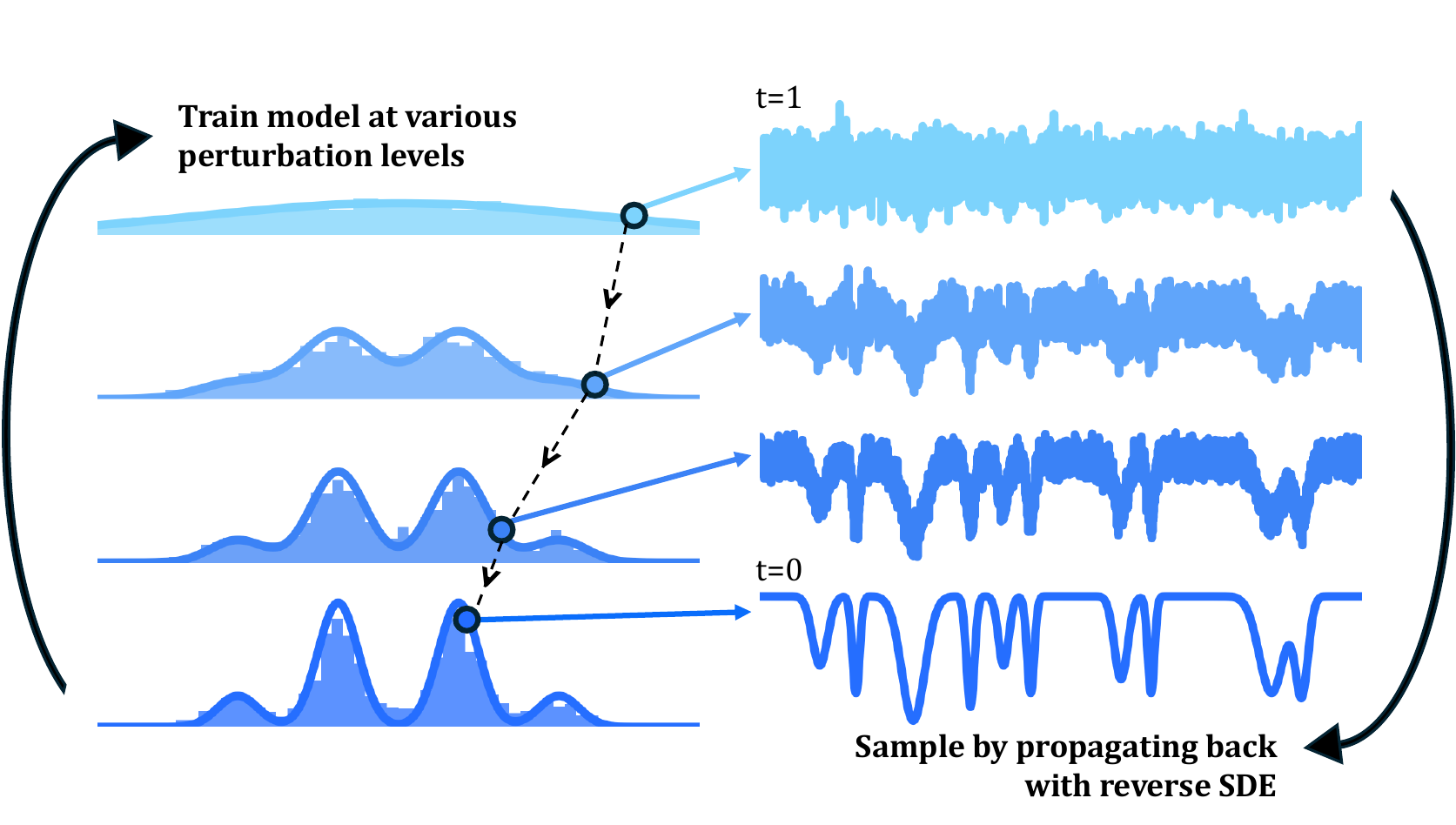}
    \caption{Schematic of the score-based diffusion model training and sampling procedure used to generate prior spectral samples. The left panel illustrates the perturbing process applied to the prior distribution of spectra. Solid curves represent the spectral distribution at different annealing times, $t$, where $t=0$ corresponds to the unperturbed prior and $t=1$ corresponds to the fully annealed distribution, which approaches an isotropic Gaussian. The model is trained to learn the gradients of these intermediate distributions. Sampling begins by drawing a noisy spectrum from the $t=1$ distribution (top light blue point). This sample is then progressively denoised by evolving it toward lower noise levels using the learnt prior gradient and the reverse-time SDE (Equation~\ref{eq:revsde}). The histograms in the left panel show the distribution of samples obtained through this reverse sampling process and demonstrate close agreement with the target distribution. The resulting spectral samples constitute draws from an approximation to the true prior spectral distribution.}
    \label{fig:diffusion}
\end{figure*}

The complete training and sampling workflow is illustrated in Figure~\ref{fig:diffusion}. We first train the SBM to learn the score of the log prior across perturbation levels. We then sample from the highest perturbation level, corresponding to nearly pure Gaussian noise, and propagate these samples backward using the reverse SDE. The reverse SDE, guided by the SBM-estimated score, transforms the samples toward the true distribution. In this way, combining SBMs with reverse-time SDE sampling allows us to obtain samples that reliably approximate the true prior distribution of stellar spectra.

\subsubsection{Posterior Sampling}
\label{sec:postspecsample}
While the previous section describes how to sample from the prior distribution of spectra, our ultimate goal is to sample from the posterior distribution, given observations. To do this, we replace the prior term, $\nabla_{f_t}\log{p(f_t)}$ in Equation~\ref{eq:revsde} with the perturbed posterior $\nabla_{f_t}\log{p(f_t \mid Y,v)}$, which is given by, 
\begin{equation}
\begin{aligned}
   \nabla_{f_t}\log{p(f_t \mid Y,v)} =  &\nabla_{f_t}\log{p(Y \mid f_t,v)}\\
   &+ \nabla_{f_t}\log{p(f_t)}.
\end{aligned}
\label{eq:posterior}
\end{equation}
Here, the annealed prior term is approximated using our trained SBM. The annealed likelihood is, in general, intractable to compute, and there exist different techniques and approximations that perform differently depending in the particular inverse problem at hand~\citep{zheng2025inversebench}. In this work, we use the Convolved Likelihood Approximation (CLA), whose full derivation is provided in Appendix~\ref{app:cla}~\citep{Adam2022}. The annealed likelihood term takes the form
\begin{equation}
    p(\mu(t)Y \mid f_t,v) = \mathcal{N}(\mu(t)Y \mid \tilde{A}f_t,\mu^2(t)\Sigma+\sigma^2(t)\tilde{A}\tilde{A}^T),
\end{equation}
where $\tilde{A}$ encodes the forward model for the spectrum and $\Sigma$ is a diagonal matrix that represents the observational uncertainty on $Y$. While this equation expresses the likelihood in terms of the scaled observation $\mu(t)Y$, the score of this expression is the same as for $\nabla_{f_t}\log{p(Y \mid f_t,v)}$ since $\mu(t)$ is only dependent on $t$ and not on $f_t$. 

To compute $\tilde{A}$, we apply the forward model in Equation~\ref{eq:forwardmodel} to a random spectrum sample $f$ and calculate the Jacobian with respect to $f$ using \texttt{torch.autograd.functional.jacobian}. While $\tilde{A}$ is invariant to the choice of $f$, it depends on the observation's radial velocity. Since the radial velocity is kept fixed in the spectrum sampling process, we only need to calculate $\tilde{A}$ once at the beginning of the sampling process. While the planetary RV's impact is negligible, a distinct $\tilde{A}$ must be generated for each observation with a different $v_{\text{BERV}}$.

Given a $f_t$ and perturbation level $t$ during the reverse SDE, the gradient of the annealed likelihood is, 
\begin{equation}
\begin{aligned}
\nabla_{f_t}\log p(Y \mid f_t, v)
&= \nabla_{f_t}
\sum \frac{-1}{2}
\Big[
\big[\mu(t)Y - A(f_t,v)\big]^{\mathsf T} \\
&\qquad \times
\big[\mu^2(t)\Sigma + \sigma^2(t)\tilde{A}\tilde{A}^{\mathsf T}\big]^{-1}\\
&\qquad \times\big[\mu(t)Y - A(f_t,v)\big]
\Big].
\end{aligned}
\label{eq:llk}
\end{equation}

The inverse of the uncertainty matrix is computed using the Cholesky decomposition to efficiently handle its large size. The gradient with respect to $f_t$ is computed using \texttt{torch.func.grad}. To avoid potential extrapolation errors from Doppler shifting, we omit the outer $1\%$ of the spectrum in the likelihood calculation. The summation is computed over all wavelength bins in $\lambda_{\text{obs}}$ and over all $N_{\mathrm{obs}}$ observations. Thus, the spectral content of all observations informs the posterior spectrum sample. 

In summary, we develop a method, \texttt{postellar}, where posterior spectrum sampling is performed using the reverse SDE (Equation~\ref{eq:revsde}), with the annealed prior replaced by the annealed posterior (Equation~\ref{eq:posterior}). The annealed prior is obtained from an SBM, $s_{\theta}(f_0,t)$, and the annealed likelihood is calculated via Equation~\ref{eq:llk}. With a trained SBM using denoising score matching and a linearizable forward model to compute $\tilde{A}$, \texttt{postellar} enables sampling from the approximate posterior distribution of stellar spectra given the data.

\subsubsection{Our Training Data}
In the previous sections, we described how an SBM is trained on a dataset of spectra to learn the prior distribution. The training spectra must be free of observational noise while still capturing the structure of real stellar spectra, so that the model does not learn noise artifacts and incorporate them into the prior. We therefore construct the training set from PHOENIX model spectra~\citep{Phoenix}. This approach is similar to previous methods that begin with stellar atmospheric models and subsequently refine or interpolate them using the likelihood~\citep{rajpurohit_exploring_2018,gully-santiago_interpretable_2022}. However, by using an SBM, we obtain a significantly more flexible and adaptive prior that is learned directly from the model spectra.
 
We focus on M-dwarf stars because they are key hosts in the search for habitable exoplanets~\citep{charbonneau_dynamics-based_2007,DressingCharbonneau}. Their low masses produce larger radial velocity signals for a given planetary mass, while their dense molecular absorption features provide a high RV information content. However, these molecular bands are notoriously difficult to model due to incomplete line lists, making M-dwarf spectra among the most challenging to reproduce.

This makes M dwarfs an ideal test case for this method, allowing us to assess whether priors learned from imperfect stellar models can successfully recover empirical spectra. M-dwarf analysis also stands to benefit the most from this approach, as it provides an alternative to both stellar atmosphere models and empirical templates. However, it should be noted that this technique can apply to any spectral type.

Since M dwarfs emit most of their flux in the near-infrared, where many of these molecular absorption bands are located, we focus on a spectrograph operating primarily in this wavelength regime~\citep{artigau_optical_2018}. We base our model on SPIRou (SPectropolarimètre InfraRouge), a near-infrared spectropolarimeter at the Canada-France-Hawai'i Telescope in Maunakea, Hawai'i~\citep{donati_spirou_2020}. SPIRou is an echelle spectrograph with $49$ orders spanning a wavelength range of about $960-2500\,\text{nm}$.

We use high-resolution PHOENIX spectra as the training set for the SBM. In the near-infrared, up to $2500\,\mathrm{nm}$, the PHOENIX models have a native sampling resolution of $R_{\mathrm{native}} \approx 500{,}000$, corresponding to the direct output of the PHOENIX code.

We use M-dwarf spectra with effective temperatures between $2300$ and $4000\,\mathrm{K}$, spanning the full range of available metallicities and surface gravities. Each spectrum is interpolated onto a finer intrinsic wavelength grid, $\lambda_{\mathrm{int}}$, with a sampling resolution of $R_{\mathrm{sample}} \approx 600{,}000$ using cubic splines with scipy's \texttt{InterpolatedUnivariateSpline} function~\citep{2020SciPy-NMeth}. This upsampling step reduces interpolation errors that could otherwise bias the RV retrieval.

The spectra are normalized by their per-order median flux and convolved with a Gaussian kernel, $\sigma_{\mathrm{kernel}} = \frac{1}{2\kappa_{\mathrm{obs}}\sqrt{2\ln{2}}}$ with $\kappa_{\mathrm{obs}} = 70{,}000$, to match SPIRou’s spectral resolution, consistent with the assumptions in Equation~\ref{eq:data}. The spectra are then padded so that their lengths are divisible by $32$, as required by the downsampling operations in the U-net architecture.

The spectra are randomly shuffled, with $20\%$ reserved for validation, leaving approximately $1500$ spectra for training. Owing to the limited training set size, we use a compact SBM architecture and restrict training to $6000$ optimization steps to mitigate overfitting.

Finally, we train a separate SBM for each SPIRou echelle order and perform RV and spectrum sampling independently for each order. This modular approach enables the exclusion of problematic orders and reduces memory demands by allowing efficient parallelization.

\subsection{Radial Velocity Sampling Framework}
\label{sec:rvsampling}

We develop an RV retrieval technique based on the Metropolis-adjusted Langevin algorithm (MALA). In this framework, the stellar spectrum is treated as fixed, and we sample the conditional posterior distribution $\log p(v \mid Y, f)$. While spectra drawn from the posterior spectrum sampler using the SBM can be combined with any RV inference method, we adopt MALA to obtain full posterior samples of the RV and thus robust uncertainty estimates. Additionally, MALA is well suited for parallelization across multiple spectra and observations and is compatible with GPU-based computation.

In MALA, proposed RV steps are generated using Langevin dynamics, which combine the score of the log posterior with a stochastic Brownian motion term, and are accepted or rejected using a Metropolis--Hastings criterion. We assume a uniform prior on $v$, such that the posterior is proportional to the likelihood, $p(Y \mid f, v) = \mathcal{N}\!\left(Y \mid A(f,v), \Sigma\right)$. As in the previous section, we omit the outermost $1\%$ of the spectrum in the likelihood calculation to mitigate extrapolation artifacts introduced by Doppler shifting in the forward model.

We define $v$ to include only the RV induced by the planetary signal. The BERV and the systemic velocity are treated as fixed and incorporated into the forward model, such that the total Doppler shift is applied once using the combined velocity. We neglect the transverse component of Earth's motion relative to the observed star, which is typically unavailable in observational data. While this approximation may introduce small inaccuracies, it ensures consistency with standard RV pipelines~\citep{wright_barycentric_2014,wright_1d_2019}.

The sampler is initialized with a given RV value, $v_0$, and a proposed step is given by, 
\begin{equation}
    v_{n+1} = v_n + s\nabla_v\log{p(Y \mid f,v_n)}+z\sqrt{2s},
    \end{equation}
where $v_n$ is the current sample, $v_{n+1}$ is the proposed step, $s$ is the predetermined stepsize, and $z\sim\mathcal{N}(0,1)$. 

The log of the acceptance ratio is then given by $\log\alpha=\mathrm{min}(0,A_r)$, where $A_r$ is given by
\begin{equation}
\begin{aligned}
    A_r = &\log p(Y \mid f,v_{n+1}) +q(v_n,v_{n+1}) \\
    &-\log p(Y \mid f,v_n) -q(v_{n+1},v_n),
\end{aligned}
\end{equation}
where
\begin{equation}
    q(a,b) = -0.5\left[\frac{(a-b-s\nabla_v\log{p(Y \mid f,b)})^2}{2s}\right].
\end{equation}

We then generate a uniform random number $u\sim\log(\mathcal{U}(0,1))$ and accept the proposed step if $u\leq\log\alpha$ and reject otherwise. 

We initialize the MALA step size, $s$, using the Bouchy uncertainty of the observation~\citep{bouchy_fundamental_2001}. An adaptive step-size scheme is then employed, adjusting $s$ until the sampler achieves an acceptance rate between $30\%$ and $40\%$. Samples generated during this adaptation phase are discarded to allow the sampler to stabilize.

This approach produces samples from the conditional posterior distribution $p(v \mid Y,f)$, enabling RV inference with any chosen spectral model. The method is GPU-compatible and readily parallelizable, allowing RVs to be inferred simultaneously for multiple observations and multiple spectral realizations.

\subsection{Initialization}
\label{sec:joint}
We initialize this procedure with the spectrum sampler, which requires fixing the RVs in order to sample the conditional posterior $\log p(f \mid Y,v)$. To obtain these initial RV estimates, we adopt a traditional template-matching approach.

The template is constructed by first interpolating all observations onto the intrinsic wavelength grid, $\lambda_{\text{int}}$. The spectra are then BERV-registered and combined by taking the median flux at each wavelength bin. Our template-matching RV estimation closely follows the classical template-matching procedure described in~\cite{silva_novel_2022}. Specifically, we estimate the RV and its uncertainty using their Equation~3, which fits a parabola to the $\chi^2$ surface using the minimum and its neighbouring points. We adopt an RV grid spacing of $1\,\text{m/s}$ for SPIRou. The resulting $\chi^2$ function is minimized using the Brent method~\citep{brent1973algorithms} and is given by
\begin{equation}
     \chi^2 = \sum_{i=1}^{N_{\mathrm{pixels}}} \frac{(Y_i - A(f_i, \ v))^2}{\sigma_{Y_i}^2 }.
     \label{eq:chi2}
\end{equation}

Using the template-derived RVs, we first perform posterior spectrum sampling using \texttt{postellar} as described in Section~\ref{sec:specsample}. The resulting spectrum samples are then fixed in the RV MALA sampler, which samples the conditional posterior $p(v \mid Y,f)$, again initialized with the template RVs. The RVs obtained from MALA can subsequently be fed back into the spectrum sampler to generate updated spectrum samples.

\begin{figure}
    \centering
    \includegraphics[width=\linewidth]{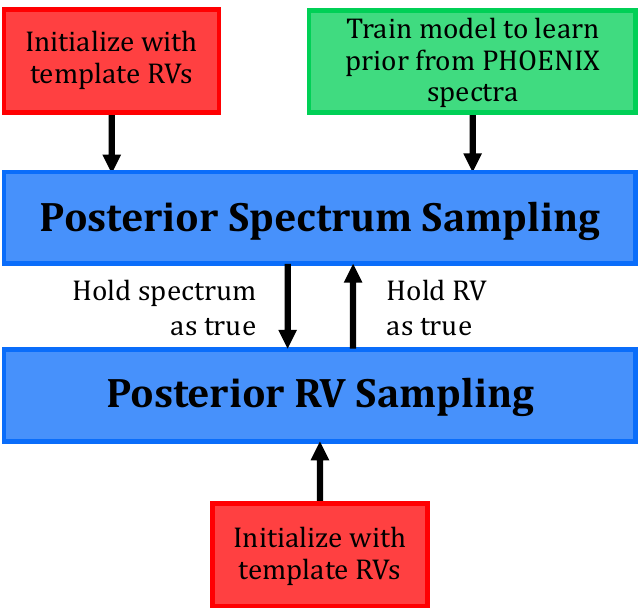}
    \caption{A flowchart demonstrating the process of sampling in which we conditionally sample the underlying spectrum and RVs from the given data, while initializing both processes with RVs retrieved using an empirical template.}
    \label{fig:process}
\end{figure}

This procedure is illustrated in Figure~\ref{fig:process}. In principle, multiple iterations are required for the Markov chain to move away from its initialization and adequately explore the joint posterior distribution of the spectrum and RV. However, since the planetary RVs considered are small and have minimal impact on the spectrum, and the initial RV value is close to the posterior mass, the results from a single step provided good empirical results. Additional iterations yielded marginal improvement in performance metrics, while significantly increasing computational cost.

\section{Results} \label{sec:results}
\subsection{Creation of Synthetic Observations}
\label{sec:syntheticobs}
To evaluate the performance of the method described above, we generate synthetic SPIRou RV observations. Synthetic data allows us to directly compare the known ground-truth spectrum and RVs with their retrieved counterparts, enabling a quantitative assessment of the method's accuracy and the reliability of the retrieved uncertainties. Moreover, constructing synthetic observations ensures consistency with the assumptions underlying Equation~\ref{eq:data}.

The synthetic observations are generated using PHOENIX spectra from our validation set, which were not used during the training process of the SBM. Because these spectra are also drawn from the PHOENIX grid, we ensure that the synthetic observations lie within the same distribution as the prior approximated by the SBM. Each native PHOENIX spectrum is interpolated onto the intrinsic wavelength grid, $\lambda_{\text{int}}$. We define the signal-to-noise ratio (S/N) as the signal-to-noise per SPIRou pixel. The spectrum is therefore normalized by its median and scaled to the appropriate flux level by multiplying by the square of the chosen S/N. We assume the underlying stellar spectrum is temporally static and neglect the effects of rotational broadening.

Because this analysis is performed on a per-echelle-order basis, the flux scaling is also applied independently to each order. As a result, the stellar spectrum has the same S/N across all orders. While this does not reflect the true stellar spectral energy distribution, it simplifies the computation and facilitates direct performance comparisons between orders. In this framework, the chosen S/N should be interpreted as the mean S/N across all orders.

Next, we define the radial velocities applied to the spectrum. Observation dates are taken to be linearly spaced over a one-year interval for a total of $N_{\mathrm{obs}}$ observations. The BERV is computed for each observation using the \texttt{helcorr} routine from the \texttt{PyAstronomy.pyasl} module~\citep{pyastronomy}. No additional systemic velocity or planet-induced radial velocity signal is included. Therefore, the ground-truth planetary RV is fixed at $0\,\mathrm{m\,s^{-1}}$ for all synthetic observations.

The synthetic spectrum is Doppler-shifted according to the predefined RVs to generate $N_{\mathrm{obs}}$ shifted spectra. The wavelength shift is computed using the relativistic Doppler equation, after which each spectrum is interpolated back onto the intrinsic wavelength grid.

These spectra are then converted into SPIRou-like observations by imposing SPIRou’s spectral resolution and sampling. Specifically, the spectra are convolved with a Gaussian kernel of width $\sigma_{\mathrm{kernel}} = \frac{1}{2\kappa_{\mathrm{obs}}\sqrt{2\ln{2}}}$, adopting a resolving power of $\kappa_{\mathrm{obs}} = 70{,}000$. Because the spectra are already expressed in terms of flux density corresponding to a given S/N per SPIRou pixel, they are subsequently interpolated directly onto SPIRou’s wavelength grid, $\lambda_{\mathrm{obs}}$.

Photon noise is then added to each wavelength bin by drawing from a Gaussian distribution with a standard deviation equal to the square root of the flux in that bin. Although photon noise is intrinsically Poissonian, we restrict our analysis to regimes in which it is well approximated by a Gaussian distribution. Other noise sources, including telluric contamination, dark current, and readout noise, are neglected to remain consistent with the assumptions of Equation~\ref{eq:data}.

To ensure consistency with the spectra used to train the SBM, the synthetic observations are normalized by dividing each spectrum by its median value as a final post-processing step.

In summary, we generate SPIRou-like observations from PHOENIX spectra that were not used during training to assess the performance of our method in recovering both the stellar spectrum and radial velocities.

\subsection{Spectrum Retrieval Performance}
We first evaluate the performance of this technique in recovering the true underlying stellar spectrum from the observations. To this end, we generate synthetic observations from a set of PHOENIX spectra spanning a range of M-dwarf spectral types. For each spectrum, we create $N_{\text{obs}} = 10$ synthetic observations at signal-to-noise ratios of $\mathrm{S/N} = 10, 50,$ and $100$. The performance is assessed across multiple echelle orders, and therefore across multiple wavelength regimes. For each set of synthetic observations, we generate five posterior samples of the spectrum.

\begin{figure}
    \centering
    \includegraphics[width=\linewidth, trim=0.3cm 1.4cm 0.3cm 3cm, clip]{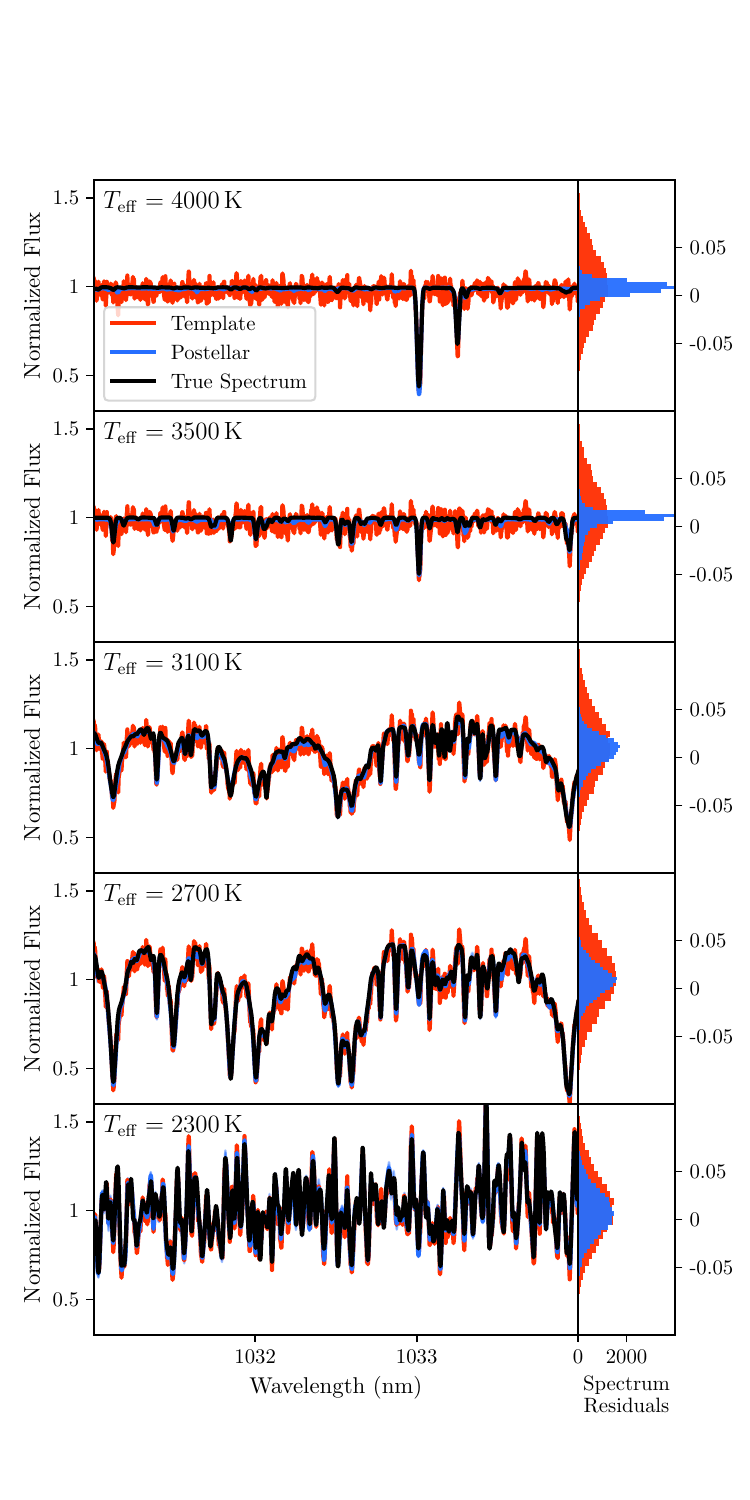}
    \caption{The black lines show synthetic SPIRou observations at an $\mathrm{S/N} = 10$  made from PHOENIX models of various $T_{\text{eff}}$. We demonstrate that \texttt{postellar}, in blue, recovers the underlying spectrum from the synthetic observations with smaller residuals than an empirical template made from the same observations, in red. The recovered spectra are shown in the left panel, and their resulting residuals are shown in the right panel.}
    \label{fig:val_phoenix_spectra}
\end{figure}

Figure~\ref{fig:val_phoenix_spectra} shows the true spectrum, the posterior spectrum samples produced by \texttt{postellar}, and a template constructed from the same synthetic observations for several M-dwarf spectra at $\mathrm{S/N} = 10$ over the wavelength range $1031$--$1034\,\mathrm{nm}$. Spectral residuals, defined as $(\mathrm{true} - \mathrm{model})/\mathrm{true}$, where the model is either the empirical template or the mean of the posterior samples, are computed over an entire echelle order spanning $1007$--$1038\,\mathrm{nm}$.

We find that the spectral residuals obtained with \texttt{postellar} are consistently narrower than those from the empirical template when compared to the true spectrum. This improvement is particularly pronounced for hotter M dwarfs. At higher stellar effective temperatures, the residual noise in the template is larger than spectral features, especially in spectrally flat regions. In contrast, \texttt{postellar} leverages its prior knowledge of the spectral structure, effectively modelling these regions as smooth continua and substantially reducing the residuals.

Some residuals exhibit a small bias at higher temperatures; however, this bias is subdominant to the observational noise. When the mean posterior spectrum is forward-modelled and compared directly to the data, no significant bias is apparent. Consequently, any small bias present in the \texttt{postellar} spectral samples is likely to be overwhelmed by observational noise and is unlikely to have a significant impact on RV measurements.

Overall, these results demonstrate that incorporating a well-characterized prior enables posterior spectral estimates that more tightly constrain the true underlying spectrum than an empirical template derived from the same observations.

\subsection{Radial Velocity Retrieval Performance}

Having established that \texttt{postellar} accurately recovers the underlying stellar spectrum, we now assess how this performance propagates to RV retrievals. To do so, we use the posterior spectral samples from the previous section to analyze $1000$ synthetic observations generated from the same underlying stellar spectrum. These observations span a 10-year baseline, each with an independent noise realization. While each observation has its own BERV, the planetary RV is fixed at $0\,\mathrm{m\,s^{-1}}$ for all cases.

For each posterior spectral sample, we generate $1000$ RV samples and discard the first $100$ as burn-in. Consequently, we obtain $4500$ RV samples across five posterior spectral samples for each observation. This analysis is restricted to a single echelle order covering the wavelength range $1007$--$1038\,\mathrm{nm}$.

For comparison, we also retrieve RVs and associated uncertainties using an empirical template and the true underlying spectrum, applying the least-squares minimization procedure described in Section~\ref{sec:joint}. We combine the posterior RV samples for a given observation across the multiple posterior spectrum samples. The median of this combined distribution is taken as the retrieved RV, and the standard deviation is adopted as the corresponding uncertainty. This approach is justified because the RV posterior distributions are well described by Gaussians.

\begin{figure*}
    \centering
    \includegraphics[width=\linewidth,trim=2cm 1.5cm 3cm 2.5cm, clip]{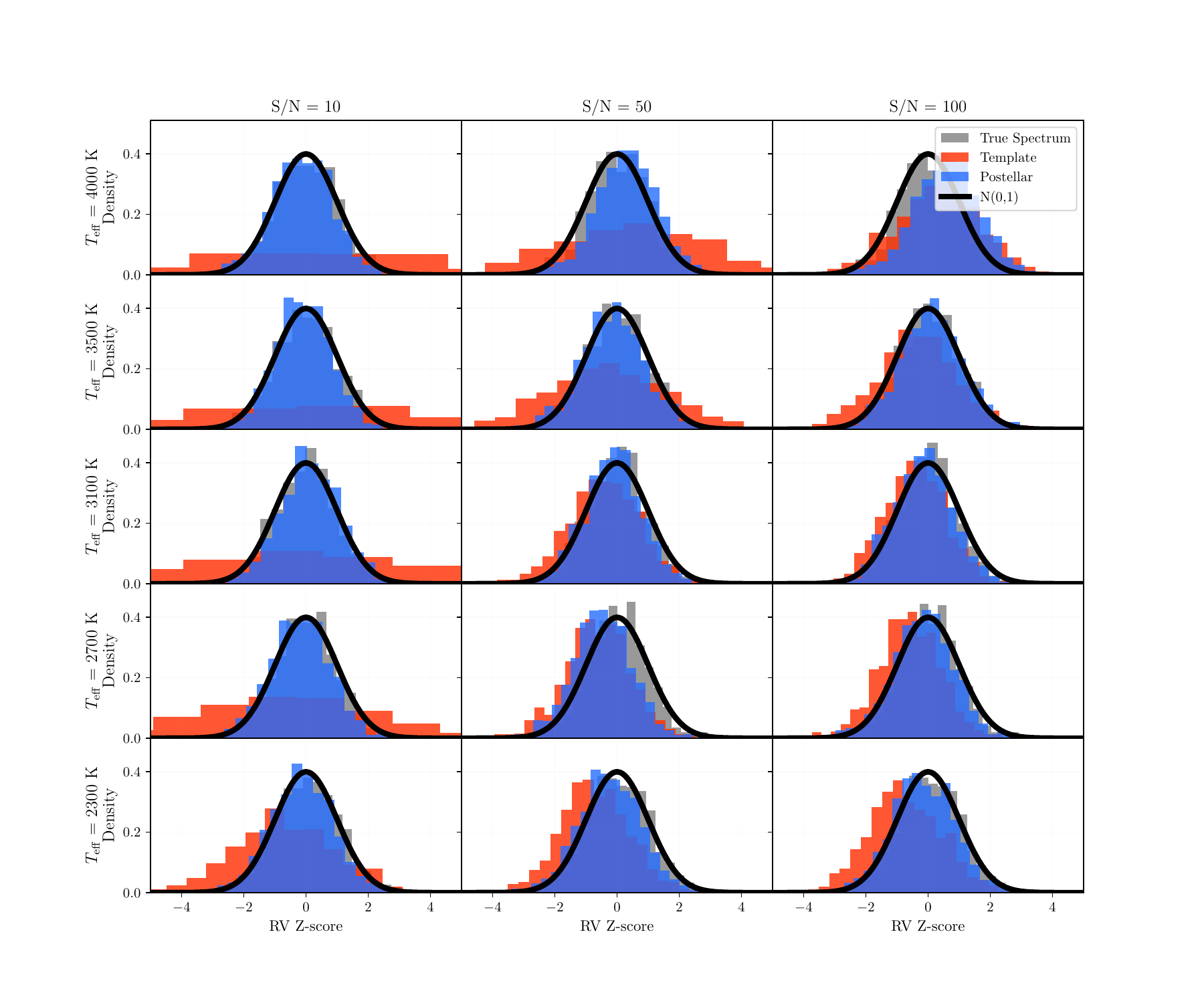}
    \caption{The Z-score distribution of retrieved RVs from synthetic SPIRou observations for various S/N regimes and PHOENIX models using the intrinsic underlying spectrum (grey), an empirical template (red), and \texttt{postellar} spectrum samples (blue). The Z-score distribution of the intrinsic spectrum and \texttt{postellar} samples largely follows a standard Gaussian distribution implying well-calibrated RV uncertainties, whereas the template Z-score distribution demonstrates under-estimated uncertainties and biased retrievals.}
    \label{fig:val_phoenix_rv}
\end{figure*}

Figure~\ref{fig:val_phoenix_rv} shows the Z-score distributions of the retrieved RVs for each S/N and spectral-type combination, comparing results obtained with the empirical template, \texttt{postellar}, and the true spectrum. The Z-score is defined as $z = (\mathrm{retrieved} - \mathrm{true})/\mathrm{uncertainty}$. If the uncertainties are accurately characterized and the retrieval is unbiased, the Z-score distribution should follow a standard normal distribution, $\mathcal{N}(0,1)$, shown by the black curve in Figure~\ref{fig:val_phoenix_rv}. When using the true spectrum, the Z-score distribution closely follows $\mathcal{N}(0,1)$, indicating unbiased RV retrievals with well-calibrated uncertainties.

We find that RVs derived using \texttt{postellar} also spectra closely follow a $\mathcal{N}(0,1)$ distribution, indicating that the associated RV uncertainties are well calibrated. In contrast, the template-matching approach substantially underestimates the scatter in the retrieved RVs, particularly at low S/N. As a result, \texttt{postellar} significantly outperforms template matching in the low-S/N regime. This behaviour is consistent with our expectation that fixing the stellar spectrum to a single template effectively collapses the joint posterior, leading to underestimated RV uncertainties.

Despite the slight spectral bias observed at higher stellar temperatures in the \texttt{postellar} reconstructions in the previous section, the corresponding RVs remain largely unbiased and continue to outperform the template-based method when uncertainties are taken into account. In fact, \texttt{postellar} performs particularly well in high-temperature regimes, where the template-derived uncertainties are severely underestimated due to the reduced RV information content of the spectra.

Overall, \texttt{postellar} produces unbiased RVs. However, in a small number of cases, typically at very high S/N or for spectra with strong RV information content, we observe a slight bias in the RV distributions. This effect may arise because spectrum sampling with the reverse SDE becomes less accurate and numerically less stable when the likelihood is extremely constraining (i.e., sharply concentrated), making the discretized reverse-SDE integration more sensitive to numerical settings~\citep{song2023pgdm, zheng2025inversebench} and potentially introducing subtle artifacts in the spectrum that bias the RV inference. 


Nevertheless, \texttt{postellar} consistently yields RV uncertainties that are well calibrated and consistent with the observed RV scatter, whereas the template-matching approach systematically underestimates uncertainties. This advantage is most pronounced at low S/N and for spectra with limited RV information content.

\subsection{Impact of Number of Observations on RV Accuracy}
In the previous section, we showed that RVs derived from \texttt{postellar} spectra yield more reliable uncertainties compared to the template method. Here, we extend this analysis to examine whether the accuracy of RV retrievals also improves. Additionally, while we previously evaluated performance using $N_{\text{obs}} = 10$ observations, the performance of template-based RV retrieval depends strongly on the number of observations used to construct the template, since the residual Gaussian noise decreases proportionally to $1/\sqrt{N_{\text{obs}}}$.

To investigate how RV accuracy depends on the number of observations informing the stellar model, we select a single PHOENIX spectrum corresponding to $T_{\text{eff}} = 3100\,\mathrm{K}$ from the previous analysis. For each chosen $N_{\text{obs}}$, we generate ten independent sets of synthetic observations, each with a unique noise realization, to sample the impact of observational noise on the spectral model. We also test a range of S/N values. For each instance, we generate a posterior spectrum sample and then retrieve RVs for $500$ synthetic observations made from the same underlying spectrum. This analysis is restricted to a single echelle order spanning $1007$--$1038\,\mathrm{nm}$.

For comparison, we also compute the retrieved RV for each configuration using the template-matching method, both with a template constructed from the same $N_{\text{obs}}$ and with the true underlying spectrum as a benchmark.
 
\begin{figure}
    \centering
    \includegraphics[width=\linewidth,trim=0.5cm 1cm 1cm 2.5cm, clip]{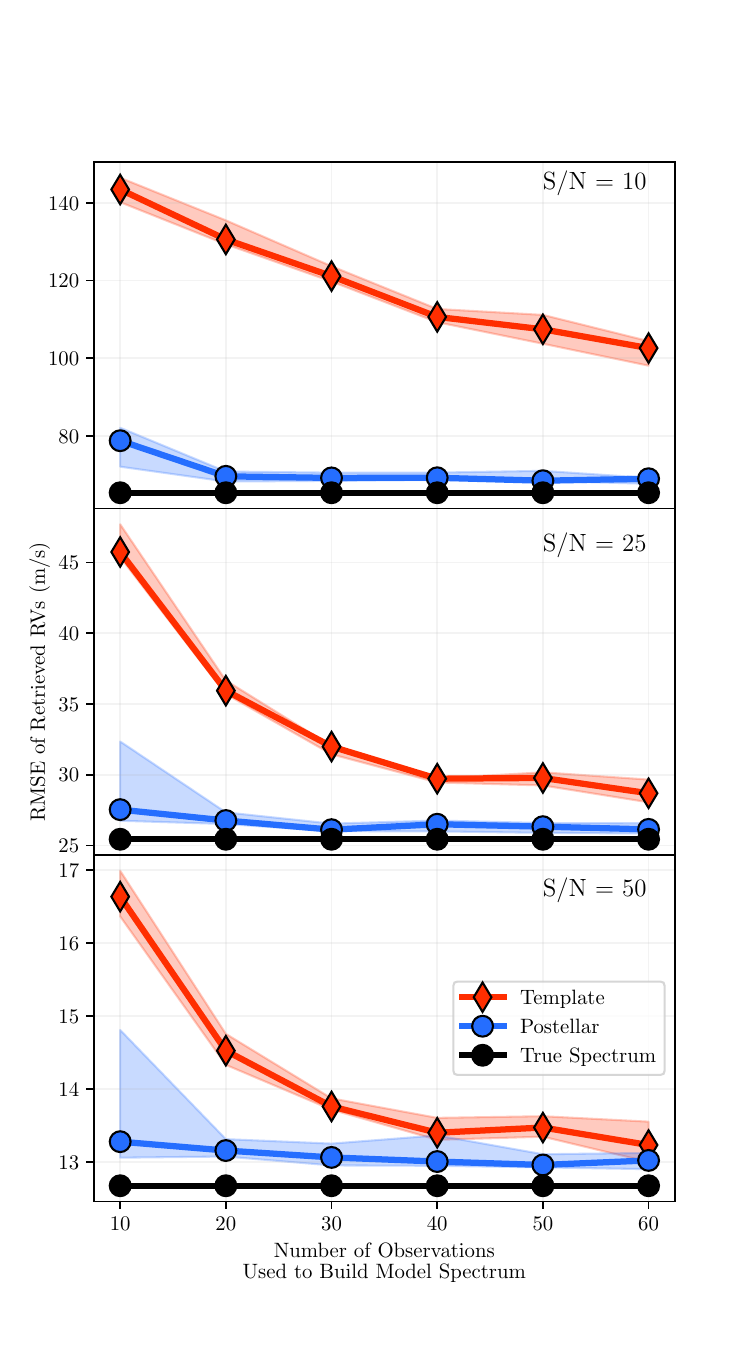}
    \caption{The median root-mean-squared error (RMSE) of retrieved radial velocities (RVs) as a function of the number of synthetic SPIRou PHOENIX-based observations used to construct the model spectrum, shown for different S/N regimes. RVs derived using the true spectrum achieve the highest accuracy and are independent of the number of observations. RVs derived from an empirical template exhibit the largest errors, which decrease as more observations are used. RVs retrieved using \texttt{postellar} spectral samples are more accurate than template-based RVs and show weak dependence on the number of observations. Shaded regions indicate the $16^{\text{th}}$--$84^{\text{th}}$ percentile range.}
    \label{fig:nobs}
\end{figure}

In Figure~\ref{fig:nobs}, we show the root-mean-squared error (RMSE) of the retrieved RVs using \texttt{postellar} spectra, the template, or the true underlying spectrum as a function of the number of observations, used to inform the stellar model. The true spectrum is independent of $N_{\text{obs}}$ and represents the best achievable RV accuracy. We find that the RV accuracy obtained with \texttt{postellar} closely approaches that of the true spectrum, even with as few as $N_{\text{obs}} = 10$ observations, across all three S/N levels.

Compared to the template method, \texttt{postellar} consistently provides better RV accuracy, particularly at low S/N and low $N_{\text{obs}}$. At low S/N, the template requires upwards of $60$ observations for its accuracy to approach that of \texttt{postellar}, while at S/N per SPIRou pixel of $50$, roughly $40$ observations are needed. Even then, the template may still underestimate the associated uncertainties.

The performance of \texttt{postellar} is largely insensitive to the number of observations used to inform the spectrum. This stability allows for selective use of observations, for example prioritizing those less affected by tellurics or stellar activity, and reduces the total observational time needed to achieve a certain level of RV performance.

Overall, using synthetic PHOENIX observations, \texttt{postellar} provides a more accurate approximation of the stellar spectrum than the template method. This translates into more accurate RV retrievals with well-calibrated uncertainties, even in regimes of low S/N or limited observational data.
 
\subsection{Application to Empirical Spectra}
We have shown that the \texttt{postellar} technique outperforms template matching at low S/N and with fewer observations when applied to synthetic spectra drawn from the same distribution approximated by the SBM. However, as noted earlier, the SBM was trained on PHOENIX models, which are known to poorly represent real M-dwarf spectra. It is therefore critical to evaluate whether \texttt{postellar} can generalize to empirical stellar spectra despite being trained solely on synthetic models.

For this evaluation, we require high-quality empirical spectra in which spectral features predominantly reflect the underlying stellar spectrum, with minimal contamination from stellar activity, photon noise, tellurics, or instrumental artifacts. We select Barnard's Star, a mid-M dwarf with $T_{\mathrm{eff}}\approx3200$K~\citep{2024A&A...690A..79G} and $J_{\rm 2MASS}=5.244$, resulting in SPIRou observations with a per-pixel S/N of $150$. Between September 2018 and October 2021, Barnard's Star was observed $871$ times as part of the SPIRou Legacy Survey, and it exhibits minimal stellar activity~\citep{donati_spirou_2020}. We therefore adopt the empirical template produced by the APERO pipeline using all SPIRou observations as its ground-truth intrinsic spectrum~\citep{cook_apero_2022}.

Similarly, Proxima Centauri, another mid-M dwarf and the closest star to the Sun ($T_{\mathrm{eff}}\approx2900$K \& $J_{\rm 2MASS} =5.357$), was observed $420$ times with the Near Infrared Planet Searcher (NIRPS) on the 3.6-m ESO telescope as part of the NIRPS commissioning program (ID 60.A-9109) and Guaranteed Time Observations (IDs 111.251P.001, 112.25NZ.001/002/003; PI: F. Bouchy)~\citep{2024SPIE13096E..0CA,mascareno_diving_2025,2025A&A...700A..10B}. We use the APERO-derived empirical template from these observations as Proxima’s intrinsic spectrum. Because NIRPS covers a narrower wavelength range (approximately $960$–$1840$ nm), only a subset of this template can be used.

To evaluate \texttt{postellar} on real stellar spectra, we create synthetic observations based on these empirical templates. First, we load the template and remove any NaN values. We then compute a wavelength grid corrected for the systemic RV, assuming $-110.47\,\text{km/s}$ for Barnard's Star and $-22.2\,\text{km/s}$ for Proxima. The template flux is interpolated onto the intrinsic wavelength grid, $\lambda_{\text{int}}$, of a given SPIRou order. Finally, the spectrum is normalized by fitting and detrending a linear model, followed by median normalization. Synthetic observations are then generated following the procedure described in Section~\ref{sec:syntheticobs}.

\subsubsection{Spectrum Reconstruction}
We create $10$ synthetic observations using the empirical templates of Proxima and Barnard's Star at various S/N values. For each case, we generate $10$ posterior samples with \texttt{postellar}, while also constructing the corresponding template by stacking the synthetic observations. The analysis is restricted to the wavelength range $1280$–$1320\,\text{nm}$.

\begin{figure*}
    \centering
    \includegraphics[width=\linewidth,trim=2.5cm 0.5cm 3cm 1.0cm,clip]{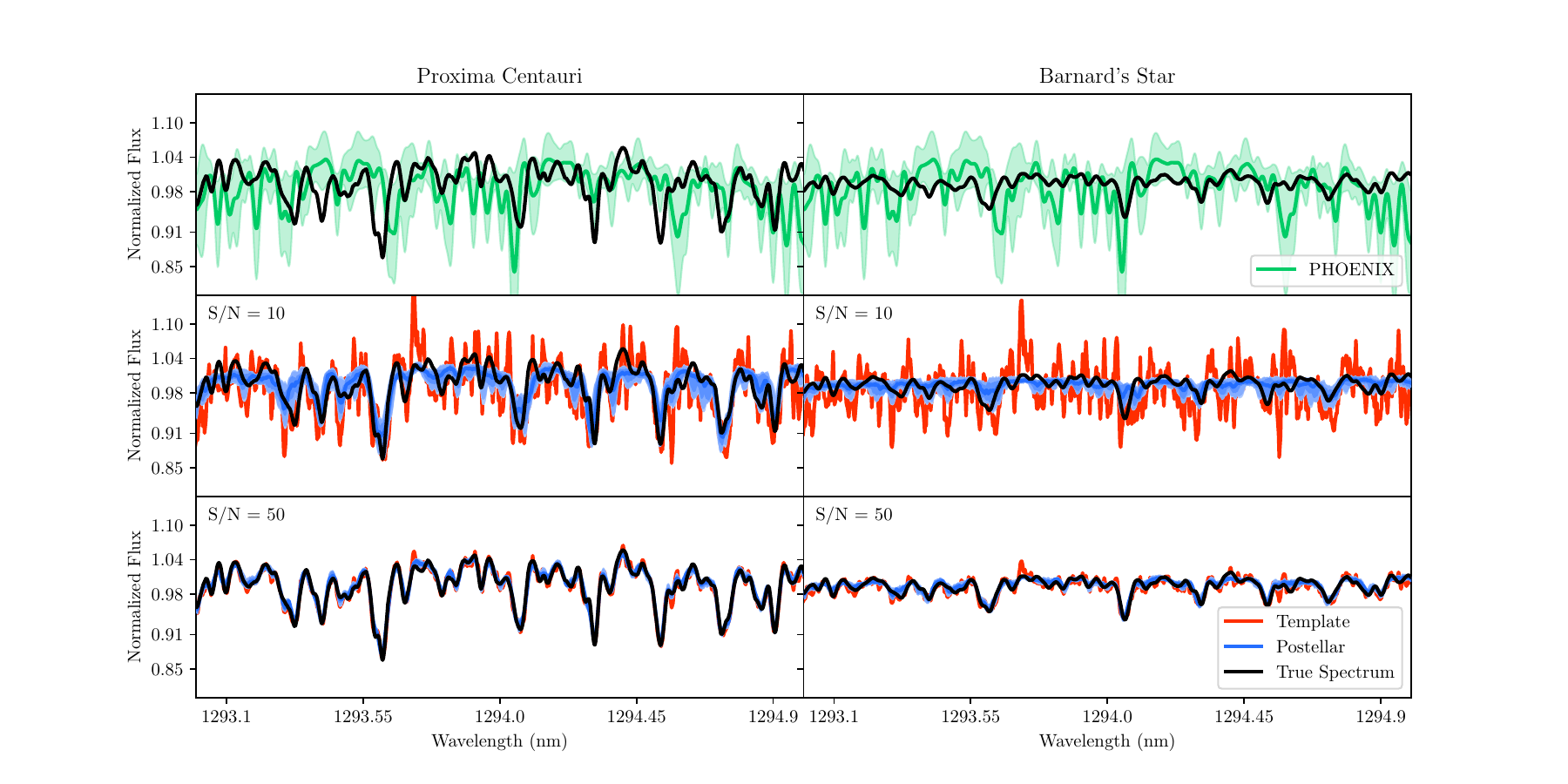}
    \caption{The black line shows the empirical high-S/N template spectrum of Proxima Centauri (left) and Barnard’s Star (right). In the top panel, the green curve shows the median spectrum obtained by taking the median flux at each wavelength across the full grid of PHOENIX M-dwarf spectra, while the shaded region denotes the corresponding $1\sigma$ envelope. Comparing this range of model spectra with the empirical M-dwarf spectra demonstrates that the PHOENIX models do not adequately reproduce the observed spectra.} The middle and bottom panels show either empirical templates or \texttt{postellar} spectral samples constructed from synthetic SPIRou observations of the respective stars at different S/N regimes. For the \texttt{postellar} spectra, the $1\sigma$, $2\sigma$, and $3\sigma$ intervals are shown, highlighting the uncertainties recovered on individual spectral features.
    \label{fig:real_spec}
\end{figure*}

The top panel of Figure~\ref{fig:real_spec} shows that the stellar spectra of Barnard's Star and Proxima are out of distribution from the PHOENIX models used to train the SBM. In the middle panel, at S/N $= 10$, \texttt{postellar} recovers the underlying spectrum more accurately than the template. The spectrum residuals are narrower than those of the template, and the posterior samples provide meaningful uncertainties. In this low-S/N regime, the prior still guides the samples, which can occasionally introduce features reminiscent of PHOENIX models, such as the slight offset of an absorption feature at $\sim 1293.2\,\text{nm}$ in Proxima. Consequently, true spectral features fall outside the 3-$\sigma$ posterior interval at a rate exceeding nominal expectations. Nevertheless, \texttt{postellar} still represents a clear improvement over the template, which provides no uncertainties.

In the bottom panel of Figure~\ref{fig:real_spec}, at higher S/N, the likelihood becomes more constraining. Here, \texttt{postellar} recovers the underlying spectrum nearly perfectly, without being misled by the out-of-distribution prior. Here, the spectral residuals remain slightly narrower than the template.

These results demonstrate that \texttt{postellar} can recover real stellar spectra even when they are out-of-distribution relative to PHOENIX models. This is possible because some features correlate between the PHOENIX models and the real spectra, allowing the method to identify these features immediately. Additionally, \texttt{postellar} leverages prior knowledge of absorption feature shapes and wavelength correlations to inform the sampling process. At low S/N, the prior guides the samples where the likelihood is weak, producing slightly broader uncertainties but still yielding lower residuals than the template. At high S/N, the method relies primarily on the likelihood to produce highly accurate and tightly constrained spectral samples.

The template is strictly limited by the photon noise in the stacked observations. In contrast, \texttt{postellar} achieves lower residuals in both low and high S/N regimes, demonstrating that incorporating prior knowledge can enable reconstruction of the stellar spectrum with uncertainties below the photon noise limit.

\subsubsection{Combined-Order Radial Velocity Analysis}
We previously evaluated the performance of \texttt{postellar} on real stellar spectra for a single echelle order. We now extend this analysis to assess its impact on RV retrievals when combining information across all usable SPIRou echelle orders. To do so, we generate synthetic observations for each SPIRou order using the empirical templates of Barnard’s star and Proxima.

The likelihood function defined in Section~\ref{sec:postspecsample} is not currently designed to handle missing data, and therefore \texttt{postellar} cannot be applied to orders containing NaNs in the empirical templates. Orders dominated by strong telluric absorption, particularly those between the $J$ and $H$ bands and between the $H$ and $K$ bands, are excluded entirely, as such regions would typically be omitted from real RV analyses. For other orders containing isolated NaNs, we assume that their RV performance is comparable to the mean performance of the analyzed orders. This approximation is necessary to construct full synthetic SPIRou datasets for Proxima.

For each usable order, we generate five posterior spectral samples using $N_{\text{obs}}=20$ synthetic observations over a range of S/N values. We then evaluate the RV performance of these spectral models using $1000$ additional synthetic observations with no injected planetary signal. Per-order RVs are retrieved for each observation, and the final RV for each observation is obtained by combining the per-order measurements using a weighted average for the template-matching technique and for the \texttt{postellar} method. This procedure yields RV “offsets” for each technique.

To assess performance in the presence of a planetary signal, we inject a circular-orbit planet with a given mass and orbital period. We simulate $20$ observations, preferentially sampling orbital phases near the extrema of the RV curve, consistent with typical follow-up strategies for transiting planets with known periods such as TESS candidates. The true RVs are generated accordingly, and the previously derived RV offsets and uncertainties are added to produce the final retrieved RV measurements. This analysis is repeated for $50$ independent observational noise realizations.

\begin{figure}
    \centering
    \includegraphics[width=\linewidth,trim=0.2cm 1cm 1cm 2cm, clip]{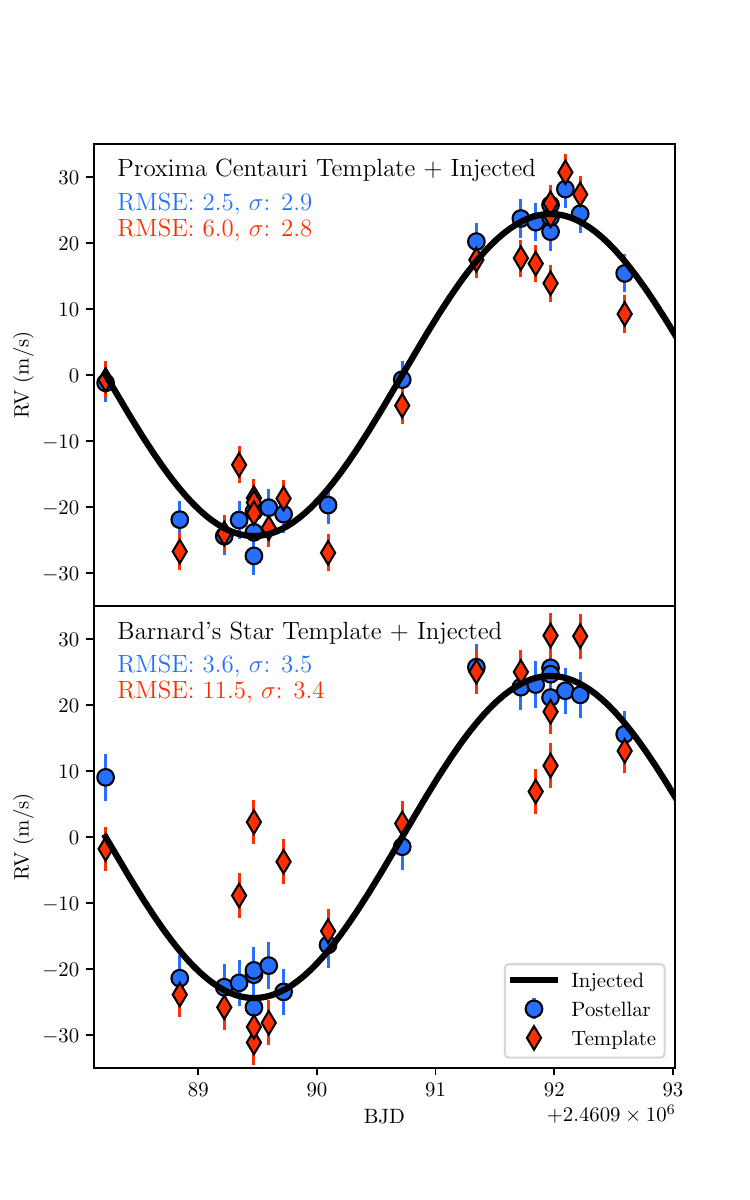}
    \caption{Retrieved radial velocities across all echelle orders for synthetic SPIRou observations generated from empirical spectra of Proxima Centauri and Barnard’s Star with an injected RV signal shown in black. RVs are derived using either \texttt{postellar} spectral samples or empirical templates constructed from $20$ observations at $\mathrm{S/N}=50$. For each case, we report the root-mean-squared error in m/s and the median RV uncertainty. RVs obtained with \texttt{postellar} are more accurate and exhibit uncertainties consistent with their scatter, whereas template-based RVs underestimate their uncertainties.}
    \label{fig:real_rvs}
\end{figure}

The results are shown in Figure~\ref{fig:real_rvs}, where we present one noise realization for both Barnard’s star and Proxima at $\text{S/N}=50$. In both cases, the \texttt{postellar} method achieves higher RV accuracy while maintaining well-calibrated uncertainties compared to the template-matching approach. At this S/N level, \texttt{postellar} reduces the RMSE of the retrieved RVs by a factor of $2$–$3$ relative to the template method. This improvement is expected to be even larger at lower S/N.

The template-matching technique consistently and substantially underestimates the RV uncertainties, even in the idealized case where photon noise is the only noise source. Although the overall RV uncertainty for Barnard’s star is larger compared to Proxima, likely due to its lower RV information content, the relative improvement provided by \texttt{postellar} is greater, as the prior more effectively constrains featureless spectral regions. This behaviour is consistently observed across all $50$ realizations.

\subsection{K-Amplitude Sensitivity Analysis}
We have shown that the \texttt{postellar} method improves RV accuracy and yields more reliable uncertainty estimates compared to template matching, particularly at low S/N and with fewer observations. We now assess how this improved RV performance translates into sensitivity to planetary signals and how it may enable more efficient observing strategies while still allowing precise planetary mass measurements.

We simulate planetary signals as described in the previous section over a range of planetary masses and orbital periods. Using the RV offsets derived from Barnard’s star posterior spectral samples constructed with $N_{\text{obs}}=20$ synthetic observations, we generate synthetic RV datasets for $50$ independent realizations of $20$ observations each. For each realization, we fit for the RV semi-amplitude, $K$, using \texttt{juliet}~\citep{2019MNRAS.490.2262E}.

For the $K$-amplitude retrieval, we adopt a uniform prior of $[-100, 100]\,\mathrm{m\,s^{-1}}$. All other orbital and instrumental parameters, such as the orbital period, transit midpoint, eccentricity, argument of periastron, and instrument offsets and jitters, are fixed to their known input values. We repeat this procedure for the $50$ realizations across a grid of planetary masses, orbital periods, and S/N values.

We find that across the S/N range tested ($\mathrm{S/N}=10$--$75$), \texttt{postellar} yields more accurate $K$-amplitude estimates with better calibrated uncertainties than the template-matching method. At higher S/N, the performance of the two methods becomes comparable. Overall, this establishes \texttt{postellar} as the preferred approach for robust planetary mass estimation, particularly in observationally limited regimes. 




As a concrete example, the sub-Neptune TOI-2136~b orbits an M4V similar to Barnard’s star and is currently the only known planet in its system. Existing SPIRou observations include $69$ measurements with an S/N of $\sim85$ in the $H$ band, yielding a $\sim3-\sigma$ detection of the RV semi-amplitude~\citep{2022MNRAS.514.4120G}. Using \texttt{postellar}, a $5-\sigma$ measurement could plausibly have been achieved with only $20$ observations at an S/N of $60$. This is enabled by the fact that TOI-2136~b is a TESS object with a well-constrained orbital period and that the stellar rotation period is significantly longer than the planetary orbital period, reducing the need for additional observations to disentangle stellar rotation signals.

Similarly, the super-Earth TOI-1452~b orbits a star of comparable spectral type and is currently the only known planet in the system. It was observed $212$ times with SPIRou at an average S/N of $\sim55$ near $1.6\,\mu\mathrm{m}$, resulting in a $\sim4-\sigma$  detection~\citep{2022AJ....164...96C}. The host star shows little evidence of stellar activity and no clear stellar rotation signal. Under these conditions, \texttt{postellar} could potentially enable a $5-\sigma$ detection with only $20$ observations at an S/N of $70$.

This improvement is particularly valuable for planet candidates around fainter M dwarfs identified by Kepler or TESS, for which orbital periods, stellar activity, and other planetary companions are known a priori but RV follow-up has been limited by target faintness. For example, the potential super-Earth TOI-1680~b orbits a star similar to Barnard’s star, that shows no clear rotational modulation or H$\alpha$ emission, and is currently the only known planet in the system~\citep{2023A&A...677A..31G}. Although its J-band magnitude of $11.6$ places it near the sensitivity limits of many northern spectrographs, SPIRou combined with \texttt{postellar} could potentially achieve a $5\sigma$ detection with $20$ observations at an S/N of $60$.

Finally, this method could also improve the efficiency of blind RV searches. For instance, GJ~1289~b, an M4-dwarf planet discovered using SPIRou RVs alone, required approximately $200$ observations at an S/N of $115$ to reach a $5\sigma$ detection~\citep{2024A&A...688A.196M}. Because the stellar rotation period is much longer than the planetary orbital period, a comparable detection could plausibly have been achieved with \texttt{postellar} using only $20$ observations at an S/N of $80$.

In practice, more than $20$ observations are typically required to robustly identify periodic signals, sample multiple orbital cycles, and account for stellar rotational variability or the presence of additional companions, particularly in the context of blind RV surveys or TESS follow-ups~\citep{cloutier_quantifying_2018}. Thus, while our analysis demonstrates that a $5-\sigma$ constraint on the $K$-amplitude can be achieved with as few as $20$ observations under idealized conditions, additional observations would generally be necessary to fully validate such a detection.

Our intent is not to suggest that such few observations are universally sufficient, but rather to emphasize that the \texttt{postellar} method enables more reliable RV measurements with substantially fewer observations than traditional template-matching approaches. In practice, this implies that robust RV signals can be identified earlier, allowing targets to be more efficiently prioritized or de-prioritized in blind surveys and TESS follow-up programs when using the \texttt{postellar} technique.

\section{Discussion}\label{sec:disc}

\subsection{Applicability to Real Observations}
The performance of \texttt{postellar} was evaluated using synthetic data to ensure that the assumptions in Equation~\ref{eq:data} were strictly satisfied and that performance could be evaluated against known ground-truth values. Here we discuss why, in spite of the assumptions made, we expect \texttt{postellar} to perform robustly on real observations. 

\subsubsection{Forward Model Misspecification}
We first assume that the instrumental line-spread function and wavelength solution are stable across observations. This assumption is well justified for SPIRou and other high-resolution spectrographs designed for precise RV measurements, where long-term instrumental stability is a core design requirement~\citep{cersullo_new_2017,hobson_spirou_2021}. In addition, we applied a model trained using SPIRou’s instrumental profile to spectra broadened with the NIRPS line-spread function and found that \texttt{postellar} still outperformed template matching. This suggests that modest mismatches or temporal variations in the instrumental profile are unlikely to significantly degrade performance.

Moreover, M-dwarf spectra are notoriously difficult to normalize due to dense molecular absorption and the absence of a well-defined continuum~\citep{rojas-ayala_metallicity_2012}. Differences in normalization between the training set and real observations could, in principle, affect performance. However, we tested \texttt{postellar} on real stellar templates that were independently normalized, and re-normalized each spectral order by its median flux to better match the training data. Given this validation, we do not expect continuum normalization to pose a significant limitation when applying \texttt{postellar} to real observations.

Our forward model also does not explicitly include additional stellar signals. Fast stellar rotation, for example, could reduce performance because the model has not been trained on spectra with strongly broadened lines. However, such stars are generally avoided in RV surveys due to their intrinsically low RV information content. In any case, this limitation could be mitigated by including rotationally broadened spectra in the training set. Other time-variable stellar phenomena, such as spots, plages, or flares are likewise not modelled. Nevertheless, because \texttt{postellar} requires fewer observations to constrain the intrinsic stellar spectrum, standard activity indicators can be used to identify and exclude contaminated observations without compromising the analysis~\citep{dumusque_measuring_2018,colliercameron_separating_2021,bellotti_mitigating_2022,simola_accounting_2022,lafarga_carmenes_2023,larue_chromaticity_2025,charpentier_reliable_2025}.

\subsubsection{Noise Model Misspecification}
We also assume that photon noise dominates and follows a Gaussian distribution. This assumption is reasonable for RV observations and for data that have already been corrected for major systematics such as tellurics and instrumental effects~\citep{bertaux_tapas_2014,gullikson_correcting_2014,artigau_telluric-line_2014,smette_molecfit_2015,leet_toward_2019,cretignier_yarara_2021,allart_automatic_2022}. Residual noise sources, such as microtelllurics, will remain, but these affect all RV extraction methods similarly~\citep{cunha_impact_2014,wang_characterizing_2022}. The Gaussian approximation may break down in low S/N regimes where readout noise dominates and photon counts are Poissonian. \texttt{postellar} is therefore intended for use in noise regimes where the Gaussian likelihood is appropriate, but future works could include learning the non-Gaussian noise to include in the sampling process~\citep{slic}.

Classical template-matching approaches rely on many of the same assumptions as \texttt{postellar}. Templates are typically constructed under the assumption that contributing observations are free of stellar activity and telluric contamination, yet they are often made from data that violate these conditions. Moreover, templates are derived from discretely sampled spectra, introducing interpolation artifacts that can propagate into RV measurements~\citep{2025AJ....170..269D}. 

In contrast, \texttt{postellar} produces a continuous spectral representation informed by a prior, rather than being determined solely by the likelihood. As a result, it should perform at least as well as template matching, and in many cases better since the prior provides physics-informed constraining power. Importantly, because \texttt{postellar} yields posterior uncertainties on the stellar spectrum, regions affected by stellar activity or tellurics manifest themselves as increased uncertainty rather than contributing to biased RV estimates.

\subsection{Limitations}
\subsubsection{Computational Considerations}
While the \texttt{postellar} framework yields stellar spectrum samples that more tightly constrain the intrinsic stellar spectrum than empirical templates, it comes at a higher computational cost. The method requires training a separate SBM for each echelle order of a given instrument. For SPIRou, this corresponds to roughly $50$ orders, each spanning $\sim30\,\mathrm{nm}$. Training a single order required a wall-clock time of approximately 7 hours on an NVIDIA A100 GPU. Importantly, though, this training step is performed only once per instrument and does not need to be repeated for each dataset.

However, the posterior sampling stage is also computationally intensive. Generating five posterior samples in parallel that are informed from $20$ observations for a single echelle order required roughly $20$ minutes on an NVIDIA H100 GPU, with a peak memory usage of $\sim60$ GB. The memory cost arises from storing $\tilde{A}\tilde{A}^{T}$ for each observation, where each matrix has dimensions on the order of $\mathbb{R}^{5{,}000 \times 5{,}000}$. The computational cost is consequently dominated by the inversion of these large matrices for the likelihood calculation.

Although this is currently GPU-intensive, the $\tilde{A}\tilde{A}^{T}$ matrices are sparse and tridiagonal. This suggests that future implementations leveraging sparse matrix representation with KeOps or other GPU-optimized libraries could substantially reduce both memory and computational requirements, making \texttt{postellar} more practical for large datasets~\citep{JMLR:v22:20-275}.

\subsubsection{Mitigating Stellar Prior Bias}
Despite decades of development, current stellar atmosphere and spectral synthesis models struggle to reproduce observed spectra of cool stars across all wavelengths. Discrepancies arise from incomplete molecular opacities, simplified physical assumptions, and empirical evidence from high-resolution spectroscopy demonstrates that our current stellar models can entirely misrepresent lines~\citep{artigau_optical_2018,Jahandar2024,Iyer2025,Glidden2026}. Since our prior is built from these incomplete models, it can be misspecified, which may bias posterior sampling. This is evident in our results as the method performs excellently on spectra within the PHOENIX model distribution but struggles more with real empirical spectra.

To mitigate prior misspecification, several approaches are possible. One is to train the model using spectra from multiple other stellar models such as from BT-Settl~\citep{BT-settl}, MARCs~\citep{Marcs2008}, or SPHINX~\citep{Iyer2025}. However, there still remain inconsistencies within these models demonstrating the need for more high-resolution models that incorporate 3D structure, non-LTE effects, and updated molecular opacities.

Alternatively, high-calibre empirical stellar libraries could supplement the training set, similar to APOGEE or IRTF but at high resolution~\citep{irtf,apogee}. Examples can include the high-quality M-dwarf spectra from CARMENES~\citep{carmenessurvey} or the gr8stars catalougue in the optical~\citep{gr8stars}. Another option is iterative prior updating where posterior samples from observed spectra, such as the samples produced from synthetic observations of Barnard’s Star and Proxima, could be added to the training dataset~\citep{Barco2025}.

In summary, avoiding prior misspecification requires both better stellar models and high-resolution empirical spectral libraries, ideally combined with iterative updates to the prior to better match real observations.

\subsection{Other Applications}
The \texttt{postellar} method provides a way to sample the underlying high-resolution stellar spectrum, which can benefit a wide range of high-resolution exoplanetary studies.

The framework can be readily extended to spectrographs other than SPIRou. Models can be retrained on PHOENIX spectra to match the wavelength solution of different spectrographs, including NIRPS, HARPS, ESPRESSO, and CARMENES. SPIRou has a relatively low sampling resolution, resulting in comparatively high S/N per pixel. Consequently, \texttt{postellar} may perform even better on higher-resolution spectrographs, provided that the noise can be approximated as Gaussian. Higher resolution also allows the prior to more accurately constrain the shapes of spectral lines. Furthermore, in the optical regime, where stellar atmosphere models are generally more accurate than in the near-infrared, the priors learned by \texttt{postellar} could be more informative during the spectral inference stage~\citep{2000ApJ...540.1005A}.

This approach is not limited to M dwarfs. In Figure~\ref{fig:val_phoenix_spectra}, \texttt{postellar} produced better residuals for higher-temperature M dwarfs because it more effectively constrains featureless spectral regions. For FGK stars, which have lower RV content but well-defined absorption lines, \texttt{postellar} could outperform template methods, as the prior provides strong guidance on line shapes and featureless regions. This simply requires training models on FGK PHOENIX spectra.

The spectrum sampling process can be further isolated and integrated with other RV retrieval techniques that traditionally rely on empirical templates. For instance, line-by-line RV analysis from ~\cite{artigau_line-by-line_2022} could be performed using the posterior spectrum samples generated by \texttt{postellar}. This approach allows for a principled way to identify and exclude spectral lines that may be contaminated by microtellurics, stellar activity, or other sources of noise, improving the robustness and precision of RV measurements. Additionally, the posterior samples provide uncertainty estimates for each spectral region, enabling more informed weighting of lines in RV calculations. 

Beyond planetary RV retrieval, the posterior spectrum samples can be applied to other RV-based analyses. For example, in Rossiter–McLaughlin measurements, which probe the spin–orbit alignment of exoplanetary systems, having precise and flexible stellar spectra can reduce systematic errors~\citep{Rossiter,McLaughlin}. Similarly, in asteroseismology studies, where tiny variations in line profiles encode stellar oscillation signals, the ability to sample the underlying spectrum with quantified uncertainties can enhance the extraction of these subtle signals and improve modelling of stellar interiors~\citep{astroseis}. The \texttt{postellar} technique may be even better for these analyses as they would not be affected by short-BERV range systematic errors that arise from templates in this regime~\citep{2025A&A...700A..93S}.

Beyond RV analysis, \texttt{postellar} samples can support other high-resolution applications. In exoplanet atmospheric characterization with transmission spectroscopy, the stellar contribution is typically removed by stacking out-of-transit observations, similar to an empirical template~\citep{snellen}. High-resolution reflected light studies also require accurate stellar subtraction, where residuals leftover from empirical templates can bias results~\citep{borra, scandriato}. Forward-modeling approaches that sample the stellar spectrum, such as that proposed by~\cite{Piskunov}, offer an alternative. In the same vein, \texttt{postellar} can provide a flexible, data-driven model of the stellar spectrum that captures uncertainties and potentially reduces residual contamination in these applications.

Moreover, in stellar abundance analysis, high S/N observations are typically required to achieve precise abundance measurements. Observations are often stacked to create a higher S/N spectrum, similar to how templates are used in precise RV analysis~\citep{kolecki}. With \texttt{postellar}, however, we can generate extremely precise stellar spectrum samples even from lower S/N observations. Abundance analysis can then be performed on the mean spectrum and its associated uncertainties using frameworks such as MOOG~\citep{sneden_moog_2012}. This approach is particularly effective for prominent atomic lines, which are already encoded in our prior and are most critical for accurate abundance determination. 

\section{Conclusion}
In conclusion, we introduce a new method, \texttt{postellar}, to infer the underlying stellar spectrum in high-resolution exoplanet spectroscopy. This framework treats the intrinsic stellar spectrum as a latent variable that can be recovered through posterior inference. We construct an informative prior over M-dwarf spectra by training a score-based diffusion model on PHOENIX stellar atmosphere models. Combined with a Gaussian likelihood, this prior enables posterior sampling through the denoising score matching technique. In this way, we coherently integrate information from physics-based models and empirical observations to recover the intrinsic stellar spectrum. Importantly, the distribution of posterior samples naturally provides uncertainties on stellar features, which can be propagated into downstream exoplanet analyses.

We apply this technique to recover radial velocities from synthetic SPIRou observations. Spectra inferred with \texttt{postellar} more accurately recover the true underlying spectrum than standard empirical templates for synthetic data generated from the empirical spectra of Barnard’s Star and Proxima Centauri. The resulting radial velocity measurements are both more accurate and accompanied by reliable uncertainty estimates, whereas velocities derived from empirical templates tend to be biased. Notably, \texttt{postellar} demonstrates substantially improved performance in the low signal-to-noise and small-sample regimes, where empirical templates are most vulnerable to noise imprinting. This makes the method particularly well suited for blind surveys and TESS follow-up observations of faint targets, where only a limited number of observations may initially be available.

Posterior inference of the stellar spectrum provides a principled framework for modelling stellar spectra while rigorously propagating stellar uncertainties into subsequent analyses. Beyond precise radial velocity measurements, this approach is broadly applicable to high-resolution studies including stellar abundance analysis, asteroseismology, and exoplanet atmospheric characterization.

\begin{acknowledgements}
We thank the anonymous referee for suggestions that greatly improved this manuscript. We would also like to thank Neil Cook and Alexandrine L'Heureux for providing helpful discussions pertaining to the NIRPS and SPIRou instruments. This work was supported by the Trottier Institute for Research on Exoplanets (IREx) and the Trottier Space Institute (TSI). D.D. is supported by the Fonds de recherche du Québec—Nature et technologies (FRQNT) (\doi{10.69777/367207}) and by the Natural Sciences and Engineering Research Council of Canada (NSERC). NBC acknowledges support from an NSERC Discovery Grant and a Tier 2 Canada Research Chair. This work was supported by grant (\doi{10.69777/377645}) from the Fonds de recherche du Québec and by the Center for research in astrophysics of Québec (AstroQuébec). The authors also thank the Trottier Space Institute (TSI) and l’Institut de recherche sur les exoplanètes (IREx) for their financial support and dynamic intellectual environment. G.M.B. acknowledges support FRQNT under a Doctoral Research Scholarship (\doi{10.69777/368273}).
\end{acknowledgements}

\appendix
\renewcommand{\thefigure}{A.\arabic{figure}}
\setcounter{figure}{0}  
\section{Convolved Likelihood Approximation}
\label{app:cla}

In this section, we will derive an analytical approximation for the annealed likelihood $p(Y \mid f_t)$, following the derivation in \citet{Adam2022} and \citet{dia2025iris}. First, let us consider that
\begin{equation}
    p(Y \mid f_t) = \frac{p(Y,f_t)}{p(f_t)}.
\end{equation}

However, since $f$ is a latent variable, the joint distribution $p(Y,f_t)$ is obtained by marginalizing over $f$. Therefore, $p(Y,f_t)$ can be written as,
\begin{equation}
    p(Y,f_t) = \int_{-\infty}^{\infty}p(Y \mid f_t,f)p(f_t \mid f)p(f)df.
\end{equation}
Since $Y$ is only dependent on $f$ (that is, $Y \perp f_t \mid f$, which implies $p(Y \mid f_t, f) = p(Y \mid f)$), we can write our annealed likelihood as such,
\begin{equation}
    p(Y \mid f_t) = \frac{\int_{-\infty}^{\infty}p(Y \mid f)p(f_t \mid f)p(f)df}{p(f_t)}.
\end{equation}

Now the Convolved Likelihood Approximation (CLA) is such that we can approximate the ratio $p(f)/p(f_t)$ as approximately constant over the region of $f$ contributing most to the integral, and absorb it into the normalization. This approximation holds true when $t\approx0$. Therefore, it converges to the true likelihood at low perturbation levels. At higher perturbation levels, it works if the likelihood is narrow in comparison to the prior. CLA has been shown to work in numerous other studies, and we demonstrate its effectiveness in our work as well. 

Our annealed likelihood can be written as,
\begin{equation}
    p(Y \mid f_t) \approx \int_{-\infty}^{\infty}p(Y \mid f)p(f_t \mid f)df.
\end{equation}

We know that the observational noise model and the perturbation of the diffusion process are Gaussian, in particular,
\begin{equation}
\begin{aligned}
    Y &= \tilde{A}f + \eta, \qquad \eta \sim \mathcal{N}(0,\Sigma), \\
    f_t &= \mu(t)f + \sigma(t)z, \qquad z \sim \mathcal{N}(0,I).
\end{aligned}
\end{equation}
This implies the conditional distributions
\begin{equation}
    p(Y \mid f) = \mathcal{N}(Y \mid \tilde{A}f,\Sigma),
    \qquad
    p(f_t \mid f) = \mathcal{N}(f_t \mid \mu(t)f,\sigma(t)^2 I).
\end{equation}
Substituting these expressions into the convolved likelihood yields
\begin{equation}
    p(Y \mid f_t)
    =
    \int_{-\infty}^{\infty}
    \mathcal{N}(Y \mid \tilde{A}f,\Sigma)
    \mathcal{N}(f_t \mid \mu(t)f,\sigma(t)^2 I)\,df.
\end{equation}

To obtain an analytical form, we consider the reparameterized residual
\begin{equation}
    \eta_t := \mu(t)Y - \tilde{A}f_t.
\end{equation}
We can expand $\eta_t$ as
\begin{equation}
\begin{aligned}
    \eta_t
    &= \mu(t)(\tilde{A}f+\eta) - \tilde{A}(\mu(t)f+\sigma(t)z) \\
    &= \mu(t)\tilde{A}f + \mu(t)\eta - \mu(t)\tilde{A}f - \sigma(t)\tilde{A}z \\
    &= \mu(t)\eta - \sigma(t)\tilde{A}z.
\end{aligned}
\end{equation}
Since $\eta$ and $z$ are independent Gaussians and $\tilde{A}$ is linear, $\eta_t$ is Gaussian with $\mathbb{E}[\eta_t] = 0$ and covariance matrix $\Sigma_{\eta_t}$ given by
\begin{equation}
\begin{aligned}
    \Sigma_{\eta_t}
    &= \mathbb{E}\!\left[\eta_t \eta_t^\top\right] \\
    &= \mathbb{E}\!\left[(\mu(t)\eta - \sigma(t)\tilde{A}z)(\mu(t)\eta - \sigma(t)\tilde{A}z)^\top\right] \\
    &= \mu(t)^2 \mathbb{E}\!\left[\eta\eta^\top\right]
       - \mu(t)\sigma(t)\tilde{A}\,\mathbb{E}\!\left[z\eta^\top\right]
       - \mu(t)\sigma(t)\mathbb{E}\!\left[\eta z^\top\right]\tilde{A}^\top
       + \sigma(t)^2 \tilde{A}\,\mathbb{E}\!\left[zz^\top\right]\tilde{A}^\top \\
    &= \mu(t)^2 \Sigma + \sigma(t)^2 \tilde{A}\tilde{A}^\top,
\end{aligned}
\end{equation}
where we used $\mathbb{E}[\eta\eta^\top]=\Sigma$, $\mathbb{E}[zz^\top]=I$, and independence of $\eta$ and $z$ (so $\mathbb{E}[z\eta^\top]=\mathbb{E}[\eta z^\top]=0$).
Therefore, under the convolved likelihood approximation, we model $\eta_t$ as
\begin{equation}
    \eta_t \sim \mathcal{N}\!\left(0,\ \mu(t)^2 \Sigma + \sigma(t)^2 \tilde{A}\tilde{A}^\top\right).
\end{equation}
Using $\eta_t = \mu(t)Y - \tilde{A}f_t$, we obtain the closed-form approximation
\begin{equation}
    p(\mu(t)Y \mid f_t)
    \approx
    \mathcal{N}\!\left(\mu(t)Y \ \middle|\ \tilde{A}f_t,\ \mu(t)^2 \Sigma + \sigma(t)^2 \tilde{A}\tilde{A}^\top\right).
\end{equation}

Because $\mu(t)$ depends only on $t$ (not on $f_t$), expressing the likelihood in terms of the scaled observation $\mu(t)Y$ rather than $Y$ changes $\log p(\cdot \mid f_t)$ only by an $f_t$-independent constant, and therefore yields the same score $\nabla_{f_t}\log p(\cdot \mid f_t)$.

\bibliography{ref}{}
\bibliographystyle{aasjournal}

\end{document}